\documentclass[11pt,preprint]{aastex}
\usepackage{emulateapj5}
\usepackage{epsfig,natbib}
\slugcomment{\rm To appear in {\it Ap. J.}, {\bf 599} (Dec. 20, 2003)}
\shortauthors{Eracleous \& Halpern}
\shorttitle{AGN Double-Peaked Emission Lines}

\def\apj{\rm {ApJ}}                

\def\cd{c$\!\!\!\hskip 0.75pt$\raise 0.2pt \hbox{\symbol{24}}}

\def\Msol{\ifmmode{\rm M}_{\mathord\odot}\else M$_{\mathord\odot}$\fi}
\def\Mbh{\ifmmode{M_{\rm bh}}\else{$M_{\rm bh}$}\fi}

\def\ls{\lower 2pt \hbox{$\;\scriptscriptstyle \buildrel<\over\sim\;$}} 
\def\gs{\lower 2pt \hbox{$\;\scriptscriptstyle \buildrel>\over\sim\;$}}

\def\kms{\ifmmode{~{\rm km~s^{-1}}}\else{~km~s$^{-1}$}\fi}
\def\m#1{\ifmmode{^{-#1}}\else{$^{-#1}$}\fi}
\def\asec{\ifmmode{^{\prime\prime}}\else{$^{\prime\prime}$}\fi}
\def\asecb{\ifmmode{^{\prime\prime\!\!\!}}\else{$^{\prime\prime\!\!\!}$}\fi}
\def\asecp{\ifmmode{^{\prime\prime\!\!\!}.}\else{$^{\prime\prime\!\!\!}$.}\fi}
\def\deg{\ifmmode{^{\circ}}\else{$^{\circ}$}\fi}
\def\degp{\ifmmode{^{\circ\!\!\!}.}\else{$^{\circ\!\!\!}$.}\fi}

\def\ten#1{$10^{#1}$} 

\newcounter{species}
\def\ion#1#2{\setcounter{species}{#2}#1$\;${\sc\roman{species}}\relax}

\def\l{$\lambda$}
\def\tl{$\lambda\lambda$}

\def\a{$\alpha$}
\def\b{$\beta$}

\def\asca{{\it ASCA}}
\def\hst{{\it HST}}

\begin{document}

\title{Completion of a Survey and Detailed Study of Double-Peaked Emission
Lines in Radio-Loud AGNs.}

\author{Michael Eracleous\altaffilmark{1,2}}
\affil{Department of Astronomy and Astrophysics, The Pennsylvania State
University, 525 Davey Lab, University Park, PA 16803}
\and

\author{Jules P. Halpern\altaffilmark{1}}
\affil{Department of Astronomy, Columbia University, 550 West 120th St., 
New York, NY 10027}

\altaffiltext{1}{Visiting astronomer, Kitt Peak National Observatory,
which is operated by the AURA, Inc., under agreement with the National
Science Foundation} 

\altaffiltext{2}{Visiting astronomer, Cerro Tololo Inter-American
Observatory, which is operated by the AURA, Inc., under agreement with
the National Science Foundation}

\begin{abstract}
We report the completion of a survey of radio-loud active galactic
nuclei (AGNs) begun in an earlier paper with the main goal
of finding and studying broad, double-peaked Balmer lines. We present
H\a\ spectra of 13 more broad-lined objects, including 3 with
double-peaked H\a\ profiles. The final sample includes 106 radio-loud
AGNs. In our final census 20\% of objects have H\a\ lines with double
peaks or twin shoulders (the ``double-peaked emitters'') and of these,
60\% (the disk-like emitters) can be fitted quite well with a model attributing
the emission to a circular, relativistic, Keplerian disk.  In four
objects where broad H\b\ and \ion{Mg}{2} lines have been observed, we
compare the profiles with models of photoionized accretion disks and
find them to be in reasonable agreement.  We reaffirm the conclusion
of paper~I that double-peaked emitters stand out among radio-loud AGNs
on the basis of a number of additional properties that they possess:
(i) an unusually large contribution of starlight to the optical
continuum around H\a, (ii) unusually large equivalent widths of
low-ionization lines ([\ion{O}{1}] and [\ion{S}{2}]), (iii) unusually
large [\ion{O}{1}]/[\ion{O}{3}] ratios, and (iv) Balmer lines which
are on average twice as broad as those of other radio-loud AGNs and
preferentially redshifted.  We consider and evaluate models for the
origin of the lines and we find accretion-disk emission to be the most
successful one because it can explain the double-peaked line profiles
and it also offers an interpretation of the additional spectroscopic
properties of these objects. We find the alternative suggestions
(binary broad-line regions, bipolar outflows, anisotropically
illuminated spherical broad-line regions) unsatisfactory because (a)
they fail direct observational tests, (b) they cannot explain all of
the unusual properties of disk-like emitters self-consistently, or (c)
in one case the physical foundations appear to be unsound.  We suggest
that double-peaked emitters and accretion-powered LINERs are the
segment of the AGN population in which the accretion rate is
considerably lower than the Eddington rate, with the consequence that
the inner accretion disk takes the form of an ion torus and the wind
that normally enshrouds the disk proper is absent. 
\end{abstract}

\section{Introduction\label{S_intro}}

In our current working picture of active galactic nuclei (AGNs) the
underlying power source is thought to be a supermassive black hole,
which is accreting matter from the host galaxy. The accretion flow
very close to the central object is believed to form an equatorial
accretion disk, by analogy with stellar accretion-powered systems such
as cataclysmic variables and low-mass X-ray binaries. The presence of
accretion disks in AGNs, although appealing from a theoretical
perspective and generally assumed, had received only limited and
indirect observational support until a few years ago, when direct
dynamical evidence became available. This evidence comprises
double-peaked line profiles characteristic of matter rotating in a
disk, much like the double-peaked emission lines of cataclysmic
variables (see, for example, Young \& Schneider 1980; Young,
Schneider, \& Shectman 1981; Marsh 1988).

A number of authors (e.g., Oke 1987; P\'erez et al. 1988; Chen,
Halpern \& Filippenko 1989; Chen \& Halpern 1989; Halpern 1990) had
proposed that the double-peaked Balmer lines found in the optical
spectra of a handful broad-line radio galaxies such as 3C~390.3,
Arp~102B, and 3C~332 originate in accretion disks around supermassive
black holes at the centers of these galaxies.  We thus carried out a
spectroscopic survey of almost 100 moderate-redshift broad-line radio
galaxies and radio loud quasars (Eracleous \& Halpern 1994, hereafter
paper I) and discovered 19 new double-peaked Balmer lines of which a
dozen conformed with a simple, kinematic accretion disk model.  We
also found that such objects stand out on the basis of additional
spectroscopic properties of their hosts. Independent dynamical
evidence for a disk-like accretion flow in the immediate vicinity of
the central black hole (within a hundred gravitational radii) has been
provided by X-ray spectroscopy of Seyfert galaxies with \asca. The
profiles of the Fe K\a\ lines of almost all Seyfert galaxies observed
with \asca\ are extremely broad and asymmetric with full widths at
zero intensity approaching a third of the speed of light (Mushotzky et
al. 1995; Tanaka et al.  1995; Nandra et al. 1997). In most cases the
line profiles can be described very well by models attributing the
emission to the inner parts of a relativistic accretion disk.

We have continued to observe radio-loud AGNs is search of
double-peaked emitters and in order to extend our tests of models for
the origin of their broad Balmer lines.  In this paper we report the
completion of the survey begun in paper I and the refined conclusions
from it. We make use of new spectroscopic data that we have obtained
as well as a great deal of information on double-peaked emitters from
recent multi-wavelength surveys. We also take into consideration some
of the variability properties of double-peaked emitters, especially
the results of reverberation of 3C~390.3.

Because emission lines from AGN accretion disks are a tool for
studying the dynamical and thermal behavior of the disk, one of our
long-term goals is to pursue such studies by monitoring the
variability of double-peaked emitters that we have found in this
survey.  Another, more lofty, goal is to use the double-peaked
emission lines (and their variability) to test dynamical models for
the line-emitting gas in AGNs, including, for example, bipolar radial
flows, and the binary black hole hypothesis. The confirmation or
rejection of {\it any} model for AGN broad-line regions would
represent progress since the dynamics of the line-emitting gas are
still poorly understood. To this end we study the general properties
of double-peaked emitters and compare them with those of the average
radio-loud AGN. We use these properties in combination with the
profiles and relative strengths of the emission lines to assess the
applicability of various proposed scenarios for the origin of the
double-peaked lines. We find that the accretion-disk interpretation of
the line profiles to be the most appealing because it provides a
self-consistent framework within which {\it all} of the properties of
the hosts can be understood.

Throughout this paper we adopt the following nomenclature: we refer to
emission lines with double peaks or twin shoulders as double-peaked
lines and to the objects that display them as double-peaked emitters.
On occasion, we distinguish the subset of double-peaked lines that can
be fitted well with a model of a relativistic, Keplerian disk (see
\S3) by referring to them as disk-like lines; their hosts are referred
to as disk-like emitters.  We adopt a Hubble constant of $H_0=70~{\rm
km~s^{-1}~Mpc^{-1}}$ and a deceleration parameter of $q_0=1/2$.

\section{Targets, Observations, and Data} 

The target selection strategy and the motivation behind it are
described in detail paper I. In summary, we chose to observe all
moderate-redshift ($z<0.4$), radio-loud AGNs for which good H\a\
spectra were not available in the literature.  The collection of
targets was supplemented with objects (listed in paper~I) whose
spectra were drawn from the literature.  In paper~I we reported the
observation of 74 broad-lined objects from the original target list,
while here we present spectra of another 10 certified broad-lined
objects.  We also report supplementary observations of 6 objects whose
H\a\ spectra were presented in paper~I.  These observations were
carried out in order to cover a broader spectral range including their
broad H\b\ lines and, if possible, also their narrow
[\ion{O}{2}]~\l3727 and broad \ion{Mg}{2}~\l2800 lines.  Finally, we
also present spectra of 3 objects with double-peaked emission lines,
which had been originally observed by other authors but had not been
studied in detail (CBS~74, Gon{\cd}alves, V\'eron, \& V\'eron-Cetty
1998; PKS~0921--213, Simpson 1994; CSO~643, Maxfield, Djorgovski, \&
Thompson 1995).  Thus, the final collection of radio-loud, broad-lined
AGNs consists of 85 objects observed by us plus 19 objects whose
spectra were drawn from the literature (the spectrum of Pictor~A used
in paper~I was taken from the literature, while in this paper we
include a spectrum obtained by us and presented in Halpern \&
Eracleous 1994). In the process of acquiring the data we also observed
12 objects which had only narrow emission lines and seven objects
whose redshifts were higher than 0.4 (see discussion in paper~I, and
below).  The final set of object is not a complete sample in a
statistical sense, but rather a representative collection of suitable
objects in AGN catalogs, {\it circa} 1991, when the target selection
was made.

The observations presented here were carried out over several
observing runs using four different telescopes and spectrographs,
namely the 4m telescope and Ritchie-Chretien (RC) spectrograph and the
2.1m telescope and GoldCam spectrograph at Kitt Peak National
Observatory, the 4m telescope and RC spectrograph at Cerro Tololo
Interamerican Observatory, the 3m Shane telescope and Kast double
spectrograph at Lick Observatory, and the 2.4m telescope at MDM
observatory. Most observations were carried out between 1993 and 1995
and a small number of them were carried out between 1997 and 2000.
The journal of observations is given in Table~\ref{Tjour}. The spectra
were taken through a narrow slit (1\asecp7--2\asecp0) in moderate
seeing conditions (1\asecp7--2\asecp5). The slit was oriented along
the parallactic angle whenever necessary to avoid differential loss of
light resulting from atmospheric refraction. The spectra were
extracted from windows of typical width 4\asec--8\asec\ along the slit
and were calibrated in a standard fashion as described in paper I. The
final spectral resolution was approximately 6~\AA\ for the spectra
taken with the 4m telescopes and 3.5--4.5~\AA\ for the spectra taken
with the other telescopes.

Our collection of newly-observed objects includes 13 AGNs with broad
Balmer lines; their H\a\ spectra are shown in Figure~\ref{Fspec}. The
remaining 3 newly-observed objects (PKS~0511--48, 3C~381, and 3C~456)
were found to have only narrow emission lines. Finally, two objects
(PKS~1355--12, and PKS~2312--319) were found to have grossly incorrect
cataloged redshifts; as a consequence, their H\a\ lines did not fall
within the observed spectral range. The spectra of the narrow-line
radio galaxies and the objects with incorrect redshifts are included
in a companion paper in the {\it Astrophysical Journal Supplement
Series}. That paper also includes a table of accurate redshifts of 
all of the objects that we have observed in our survey.

\section{Demography of Broad Emission-Line Profiles}

As in paper~I, we identify H\a\ profiles with displaced peaks or
shoulders and divide them into groups accordingly. The noteworthy
objects from this paper are 4C~31.06, Pictor~A, CBS~74, PKS~0921--213,
and CSO~643. Our final sample includes 106 radio-loud objects of which
20--23\% have double-peaked Balmer lines (depending on whether or not
we count objects whose double-peaked lines were known {\it a
priori}). Out of a total of 24 double-peaked emitters, 17 have the
blue peak stronger than the red, which has a chance probability of
0.02. Thus, there is an intrinsic preference for the blue peak being
stronger than the red, which disfavors scenarios in which a random
distribution in phase space of line-emitting ``clouds'' gives rise to 
double-peaked
line profiles by chance. Furthermore, 13 out of the 17 profiles that
have the blue peak stronger than the red can be fitted well with a
simple disk model, as we describe in later sections.

It is important to note that membership in the above groups of line
profiles may be temporary, since double-peaked line profiles are
known to vary significantly on time scales of years. The variations
can take the form of small but significant changes in the relative
strengths of the two peaks (e.g., Arp~102B; Miller \& Peterson 1990;
Newman et al. 1997), complete reversals in the symmetry of the
profile, namely the blue peak being stronger than the red at some
epochs but not at others (e.g., 3C~390.3 and 3C~332; Zheng, Veilleux,
\& Grandi 1991; Gilbert et al. 1999), and the emergence of
double-peaked profiles with subsequent
dramatic variations (e.g., Pictor~A; Halpern \& Eracleous 1994;
Sulentic et al. 1995a; Eracleous \& Halpern 1998). Therefore, the
relative numbers of blue- and red-asymmetric double-peaked profiles
could reflect the fraction of time that the profiles spend in a given
state. For the purposes of this paper, we consider all double-peaked
emitters as a group, without making a distinction based on the sense
of their asymmetry or whether or not the simplest models can fit their
profiles in detail.

To study the broad emission-line profiles quantitatively, we measured
the line widths and shifts at half maximum and at zero intensity. The
results of these measurements, are listed in Table~\ref{Tstar}. The
measured widths and shifts at zero intensity should be regarded with
caution; their error bars (given in Table~\ref{Tstar}) are an order of
magnitude larger than those on the widths and shifts at half maximum
because of the uncertainty in determining the continuum level and
hence the uncertainty in identifying the far wings of the line.  In
Figure~\ref{Fwidths} we compare the distributions of FWHM and FWZI of
double-peaked emitters with those of other objects in our entire
collection, following paper~I. The mean values and the probabilities
that corresponding distributions are drawn from the same parent
population (according to the Kolmogorov-Smirnov, or K-S, test) are
given in Table~\ref{Tcomp}. The most significant difference is in the
distributions of FWHM with double-peaked emitters having H\a\ lines
that are on average twice as broad as other objects.  The distribution
of fractional shifts among all objects in the collection is plotted in
Figure~\ref{Fshifts}, in which the double-peaked emitters are
represented by the shaded histogram bins. There is a preference for
redshifts, with a mean shift value of
$\Delta\lambda/\lambda\approx10^{-3}$, or $\Delta v \approx 300$~\kms
(the same at half maximum and at zero intensity, both for
double-peaked emitters and for other radio-loud AGNs).  In 4/5 of
double-peaked emitters the H\a\ lines show a net redshift at half
maximum. At zero intensity the error bars are quite large, leading us
to use extremely coarse bins. Within these large error bars there are
no blueshifts among double-peaked emitters.

Our findings regarding the properties of the line profiles of the
entire set of radio-loud AGNs are by no means new results.  The widths
and shifts we measure are in agreement with the results of more
systematic and detailed studies such as those of Boroson \& Green
(1992), Brotherton (1996), Corbin (1994, 1997b), Sulentic et
al. (1995b), and Marziani et al.  (1996).  It is also noteworthy that
variants of many of these results have been known for more than 20
years (see Miley \& Miller 1979 and Steiner 1981).  Our comparison of
the properties of double-peaked emitters and other objects shows that
double-peaked emitters are extreme objects, occupying the high end of
the Balmer line width and redshift distributions. We will return to
these results in \S6 where we will discuss them in
the context of scenarios of the structure of line-emitting region.

\section{Model Fits to Double-Peaked Profiles}

\subsection{H\a\ Line Profiles\label{S_fit_Ha}}

We have tried to fit H\a\ profiles with well-defined twin peaks or
pronounced shoulders (4C~31.06, Pictor~A, CBS~74, PKS~0921--213, and
CSO~643) with the relativistic Keplerian disk model of Chen \& Halpern
(1989). This model assumes that the line originates on the surface of
a circular, Keplerian disk whose axis is inclined by an angle $i$
relative to the line of sight, between radii $\xi_1$ and $\xi_2$
(expressed in units of the gravitational radius, $r_{\rm g}\equiv
G\Mbh/c^2$, where \Mbh\ is the mass of the black hole). The disk has
an axisymmetric emissivity of the form
$\epsilon\propto\xi^{-q}$. Local broadening of the line is represented
by a Gaussian rest-frame profile of velocity dispersion is $\sigma$
(this is a combination of turbulent motions in the disk and the
velocity gradient within the cells used in our numerical
integration\footnote{The effect of local broadening can be reproduced
by a model in which the outer edge of the disk is not sharp, (see Chen
\& Halpern 1989)}). The free parameters of the model are thus
$\xi_1,~\xi_2,~i,~q$ and $\sigma$. The flux illuminating the outer
disk is expected to vary as $r^{-3}$, while the the H\a\ flux emerging
from the disk is proportional to the illuminating flux for an
extremely wide range of values of the latter (see Collin-Souffrin \&
Dumont 1989 and Dumont \& Collin-Souffrin 1990a,b). Therefore, we
adopt $q=3$ as a reasonable prescription for the H\a\ emissivity of
the disk. Lower values of $q$ can result when the H\a\ emissivity
saturates at very high values of the ionizing flux.  In
Figure~\ref{Ffit}a,b,e we show the H\a\ profiles of 4C~31.06, CBS~74,
and PKS~0921--213 with the best-fitting disk models superposed. The
model parameters are generally in the same range as the those found
from fitting the double-peaked profiles of paper~I; they are
summarized in Table~\ref{Tfit}.  The goodness of the fit was judged by
eye and the error bars on the model parameters were obtained by
perturbing each parameter about its optimal value and adjusting all
other parameters until an acceptable fit was no longer possible.

The above model does not describe the H\a\ profiles of Pictor~A and
CSO~643 as well as it does those of other objects. In the case of
Pictor~A, the red peak is stronger than the blue, contrary to the
expectations of the simplest, axisymmetric disk models, while in the
case of CSO~643 an {\it ad hoc} blueshift of 1370~\kms\ appears to be
needed for a good fit. Therefore, we attempted to fit them with the
elliptical disk model of Eracleous et al. (1995), which has two
additional free parameters.  We are not suggesting that these objects
necessarily harbor eccentric accretion disks, although we do consider
this a plausible scenario. The point of this exercise is to
demonstrate that with some refinement, accretion disk models can
account for a much wider variety of line profiles. The model
parameters for these two objects (for both circular and elliptical
disk models) are included in Table~\ref{Tfit} and the fits are
compared to the data in Figure~\ref{Ffit}c,d,f,g. We discuss the
physical justification for these models further in \S\ref{S_indiv}.

\subsection{{\rm H}\b\ and {\rm Mg~II} Line Profiles}

An important prediction of the models of photoionized accretion disks
by Collin-Souffrin \& Dumont (1989) and Dumont \& Collin-Souffrin
(1990a,b) is that, because of their large densities (of order
\ten{15}~cm\m3) and column densities (of order \ten{25}~cm\m2), the
disks would emit predominantly low-ionization lines (e.g., Balmer
lines, \ion{Mg}{2}, and \ion{Fe}{2}) rather than high-ionization
lines. Thus we have tried to verify that the profiles and strengths of
the \ion{Mg}{2}\ and H\b\ lines are consistent with an accretion disk
origin using blue spectra of four objects whose \ion{Mg}{2}\ lines are
redshifted to wavelengths longer than the atmospheric cutoff (3C~17,
PKS~0340--37, PKS~1151--34, and CSO~643; $0.2<z<0.3$). To carry out a
more complete and detailed test, we have obtained UV spectra of
several objects with relatively low redshifts ($z<0.1$) with the {\it
Hubble Space Telescope (HST)}.  The results for the first of these
objects, Arp~102B, have been published (see \S5.2), while a paper
presenting our \hst\ study of several more objects is in preparation.

According to the photoionization models we have adopted, the H\b\
emissivity is proportional to $r^{-3}$ at moderate radii but it rises
steeply towards the center at small radii (approximately as
$r^{-5.1}$). The \ion{Mg}{2}\ line emissivity is also proportional to
$r^{-3}$ at moderate radii and becomes ``shallower'' at small
radii. The emissivity of {\it all} lines saturates at very high
incident fluxes (i.e., at very small disk radii); the incident flux
required to saturate H\a\ is about an order of magnitude higher than
that for H\b\ or \ion{Mg}{2}. The emissivity laws described above are
summarized in Figure~7 of Collin-Souffrin \& Dumont (1989).  Thus, we
parameterized the the disk emissivity as a broken power law of the
form
\begin{equation}
\epsilon(\xi)=\cases{ 
\epsilon_0\; \xi^{-q_1}, & 
for $\xi_1<\xi<\xi_{\rm b};$\cr
& \cr
\epsilon_0\; \xi_{\rm b}^{q_2-q_1}\; \xi^{-q_2}, & 
for $\xi_{\rm b}<\xi<\xi_2$,\cr}
\end{equation}
where $\epsilon_0$ is a constant of proportionality, $\xi_{\rm b}$ is
the break radius at which the power-law index changes, and the factor
$\xi_{\rm b}^{q_2-q_1}$ ensures that the emissivity is continuous at
$\xi_{\rm b}$.  The power-law indices were held fixed to $q_1=5.1$ and
$q_2=3$ for the H\b\ fit. For the \ion{Mg}{2}\ fit we set $q_2=3$ and
allowed $q_1$ to vary between 2 and 3 to simulate the flattening of
the emissivity profile at small radii. The broadening parameter,
$\sigma$, and of course the inclination angle of the disk, $i$, were
held fixed to the values determined by fitting the H\a\ profile in
paper~I. We also tried to keep the inner and outer radii of the
line-emitting disk close to the values determined from fitting H\a,
although this was not always possible (the parameters determined from
fits to the H\a\ profiles are more robust because the H\a\ profile is
not seriously contaminated by strong, narrow lines, and has the simplest
emissivity law). In Figure~\ref{Fprof} we show a montage of the the
H\a, H\b, and \ion{Mg}{2}\ profiles for each of the three objects,
plotted on a common velocity scale, with the best-fitting models
superposed.  The model parameters are summarized in
Table~\ref{Tfit}. In the case of 3C~17 we used an elliptical disk
model to describe the line profile (see \S\ref{S_fit_Ha}, above); the
model for the H\a\ line also describes the low-$S/N$ profile of the
weak H\b\ line.  In the case of CSO~643 we applied a blueshift of
1370\kms\ to the model profiles (see \S\ref{S_fit_Ha} and
\S\ref{S_indiv}).

We find that the model parameters expected on theoretical grounds
produce acceptable fits to the H\b\ profiles. However, the goodness of
a fit can only be judged for the blue half of the line profile because
the red side is overwhelmed by the very strong [\ion{O}{3}]~\tl4959,
5007 doublet from the narrow-line region.  As a result, we cannot
easily distinguished between a simple and a broken power-law
emissivity prescription for the H\b\ line. The observed \ion{Mg}{2}\
profiles are also consistent with a disk origin, although they appear
to come from a larger region of the disk than what we expected on
theoretical grounds. In fact, in some cases we can only obtain a lower
limit to the outer radius of the \ion{Mg}{2}-emitting region of the disk.

The relative strengths of the H\a, H\b, and \ion{Mg}{2}\ lines of the
four objects were measured by integrating the flux in the best-fitting
models and they are included in Table~\ref{Tlrat}. This table also
includes the H\a/H\b\ ratios of all other double-peaked emitters for
which the H\b\ lines have been observed and the \ion{Mg}{2}/H\b\ ratio
for Arp~102B from Halpern et al. (1996); these ratios were measured by
superposing the two profiles and scaling them so that they
matched. All line ratios have been corrected for reddening in the
interstellar medium of the Galaxy (see \S5.4). We find that the
measured H\a/H\b\ ratios are generally consistent with the predictions
of Dumont \& Collin-Souffrin (1990a), but the measured
\ion{Mg}{2}/H\b\ ratios are higher than the model predictions by a
factor of about 3. The discrepancy can be the result of one or a
combination of the following effects:

\begin{enumerate}
\item
The disk is not the only source of \ion{Mg}{2} emission; a wind
overlaying the geometrically thin disk may contribute significantly to
the observed \ion{Mg}{2} flux but not to the observed H\b\ flux. The
profile of the \ion{Mg}{2} line from the wind is likely to be narrower
than the that from the disk and single-peaked, and it could
fill in the trough between the two peaks of the disk profile. Thus the
situation would be analogous to the \ion{H}{1} lines of Arp~102B,
where there is strong evidence that the line profiles comprise
contributions from two different sources with different dynamics and
ionization structure (Halpern et al. 1996).
\item 
The ionizing flux intercepted by the disk is so low that the metal
lines are formed primarily by collisional excitation rather than
recombination, while the Balmer lines are formed almost exclusively by
recombination. This effect is discussed by Collin-Souffrin \& Dumont
(1989) and reiterated by Dumont \& Collin-Souffrin (1990a). These
authors note that such an effect is an important characteristic of
weakly-ionized media. Thus, the \ion{Mg}{2}/H\a\ ratio rises above
unity in the outer disk, with the consequence that the outer disk
makes a significant contribution to the \ion{Mg}{2} line but not to
the H\a\ line. This behavior is not captured by the simple power-law
parameterization that we have adopted.

\end{enumerate}

Our overall conclusion is that the photoionization models of
Collin-Souffrin \& Dumont (1989) and Dumont \& Collin-Souffrin
(1990a,b) fare moderately well against the observations. They provide
emissivity prescriptions that can be used to fit the observed line
profiles well, and they can predict the H\a/H\b\ ratios fairly well.
We note in closing that as more observations of \ion{Mg}{2} lines
become available, tests of these models can be refined and extended.

\section{Observational Properties of Double-Peaked Emitters}

\subsection{Optical Spectroscopic Properties and Evaluation of Extinction
\label{S_optical}}

Following paper~I, we repeat and refine the statistical comparisons
between the starlight fraction and narrow-line equivalent widths of
double-peaked emitters using the combined data from both papers.  We
also compare the distribution of the ratios of the narrow oxygen
lines, [\ion{O}{1}]/[\ion{O}{3}] and [\ion{O}{2}]/[\ion{O}{3}], of
double-peaked emitters and other radio-loud AGNs. These line ratios
serve as a diagnostic of the ionization state of the narrow-line
region. Unlike paper~I, we put all double-peaked emitters in the same
group, without distinguishing objects whose H\a\ profiles are
well-described by a disk model (see our discussion in
\S3).

In Table~\ref{Tstar} we list the {\it observed} equivalent widths of
the [\ion{O}{1}] and [\ion{S}{2}] lines and the fractional
contribution of starlight to the optical continuum around H\a\ in the
newly observed objects.  In Table~\ref{Tlrat} we list the oxygen line
ratios of double-peaked emitters and comparison objects, which were
either measured from our own spectra or drawn from a number of sources
in the literature. The effects of Galactic reddening were corrected
using the reddening measurements of Schlegel, Finkbeiner, \& Davis
(1998) and assuming the Seaton (1979) reddening law. To guard against
uncertainties in the absolute flux calibration of any individual
spectrum, line ratios were computed only if both lines could be
measured from the same spectrum. We were able to match the redshift
range of the [\ion{O}{1}]/[\ion{O}{3}] comparison sample to that if
the double-peaked emitters (i.e., $z<0.4$). In the case of the
[\ion{O}{2}]/[\ion{O}{3}] comparison sample, however, we had to extend
the redshift range to $z<0.6$ so as to build up a sizable sample from
published data.

The results of these comparisons are shown graphically, in the form of
histograms in Figures~\ref{Fstar} and \ref{Flrat}. The average values
of the relevant quantities are summarized in Table~\ref{Tcomp} along
with the probability that their distribution in double-peaked emitters
and other radio-loud AGNs were drawn from the same parent population.
The results of this comparison reaffirm the conclusions of paper~I:
the mean starlight fraction is in double-peaked emitters is three
times higher than in other radio-loud AGNs, while the mean rest-frame
equivalent widths of the [\ion{O}{1}] and [\ion{S}{2}] lines are 36\%
and 60\% higher. The oxygen line ratios suggest that the narrow-line
regions of double-peaked emitters are at a lower ionization state than
than those of typical radio-loud AGNs. The oxygen line ratios
double-peaked emitters, are, in fact, reminiscent of LINERs
(low-ionization nuclear emission regions; after Heckman 1980),
although only Arp~102B fulfills the formal definition of that
class. We discuss this connection further in our interpretation of the
properties of double-peaked emitters in \S6.1.

In addition to the properties of the narrow lines we have also studied
the properties of the broad lines of double-peaked emitters using a
similar approach.  In Table~\ref{Tlrat} we list the H\a/H\b\ and
\ion{Mg}{2}/H\b\ ratios of double-peaked emitters and comparison
objects. Measurements, reddening corrections, and construction of
comparison samples was carried out as for the narrow-line ratios
(including extension of the redshift range of the \ion{Mg}{2}/H\b\
comparison sample to $z<0.6$).  The \ion{Mg}{2}/H\b\ ratios for many
of the comparison objects were kindly provided by M.~S.~Brotherton and
B.~J.~Wills.  In Figure~\ref{Flrat} we compare the distributions of
broad-line ratios in double-peaked emitters and in other radio-loud
AGNs, while the mean values and K-S probabilities are included in
Table~\ref{Tcomp}. We find that the distributions of the H\a/H\b\ ratio are
considerably different, while those of the \ion{Mg}{2}/H\b\ ratio are quite
similar, although this is based on a small number of double-peaked
emitters. At first glance, one can interpret the larger H\a/H\b\ ratio
of double-peaked emitters as the result of reddening, which would
imply a difference in the mean extinction in the two groups of about
$\Delta A_{\rm V}\approx 0.35$. However, this interpretation would be
in conflict with the \ion{Mg}{2}/H\b\ ratios of double-peaked emitters
which are not accordingly smaller than those of the comparison
objects. Since this statement is based only a small number of
measurements of the \ion{Mg}{2}/H\b\ ratio in double-peaked emitters,
it should be checked again as more data become available. If we take
the \ion{Mg}{2}/H\b\ distributions at face value, we conclude that the
broad-line regions of disk-like emitters are not unusually
reddened. Rather, it is more plausible that the H\a/H\b\ ratios of
double-peaked emitters are enhanced by collisional excitation, which
could be a consequence of an origin of the broad Balmer and
\ion{Mg}{2} lines in the dense material of an accretion disk.

\subsection{Properties at Other Wavelengths}

All of the double-peaked emitters considered here are radio-loud.  The
radio properties of the entire set (5~GHz monochromatic luminosities,
spectral indices, and morphologies) are listed in
Table~\ref{Tradio}. Most double-peaked emitters are powerful FR~II
radio galaxies (Fanaroff \& Riley 1974) with $L_{\;\rm 5~GHz}\gs
10^{25}$~W~Hz\m1 and double-lobed radio morphologies; three objects
are associated with powerful but compact steep-spectrum radio
sources. Four objects (IRAS~0236.6--3101, 1E~0450.3--1817,
PKS~0921--213, and Arp~102B) have $L_{\;\rm 5~GHz}\ls
10^{24}$~W~Hz\m1, which is intermediate between what are commonly
considered high and low radio luminosities.

The infrared properties of double-peaked emitters are poorly
known. Only two of the nearest and best-studied objects have been
detected by {\it IRAS}: Arp~102B and 3C~390.3. In the broad-band
spectral energy distributions (SEDs) of these two objects (Chen et
al. 1989; Golombek, Miley, \& Neugebauer 1988) the usual ``UV bump''
is absent. Instead these SEDs peak in the infrared band around
25~$\mu$m. The weakness of the non-stellar continuum in the optical
spectra of all double-peaked emitters is probably a manifestation of
the absence of a ``UV bump''.

The UV spectra of double-peaked emitters can afford important tests of
models for the origin of the double-peaked lines via the strong,
high-ionization, resonance lines that they cover. The only
double-peaked emitter with a published UV spectrum from the $HST$
(i.e., with a moderate-resolution and a high signal-to-noise ratio) is
Arp~102B (Halpern et al 1996). This spectrum shows that \ion{Mg}{2}\
is the only UV line which is as broad as the Balmer lines and that the
twin peaks of the low-ionization lines do not have counterparts in the
high-ionization lines. The upper limit to the Ly\a/H\b\ ratio in the
twin peaks is 0.15. The absence of a double-peaked Ly\a\ line cannot
be the result of reddening because this would be in conflict with
the observed \ion{Mg}{2}/H\b\ and H\a/H\b\ ratios which are typical of
radio-loud AGNs (see \S5.4). We discuss the implications of this
spectrum further in \S\ref{S_disc_tests}.  We are currently in the process
of analyzing $HST$ spectra of more double-peaked emitters.

A fair fraction of double-peaked emitters are bright enough in the
X-ray band to allow their X-ray properties to be studied. Soft X-ray
fluxes of most double-peaked emitters are listed in Table~\ref{Tdisk},
which is an expanded and updated version of a similar table from
paper~I. The hard X-ray spectra of many of the nearby double-peaked
emitters have been studied with {\it ASCA} (0.5--10~keV; see, for
example, Wo\'zniak et al. 1998 and Sambruna, Eracleous, \& Mushotzky
1999) and {\it RXTE} (4--100~keV; see Eracleous, Sambruna, \&
Mushotzky 2000, and references therein). The general conclusion is
that the X-ray spectra of most double-peaked emitters do not differ
from those of typical radio-loud AGNs.

\section{Discussion}

\subsection{Interpretation of the Properties of Double-Peaked Emitters}

The accretion disk model of Chen et al. (1989) and Chen \& Halpern
(1989) can provide an explanation for the unusual properties of
double-peaked emitters.  The predicted line profiles can fit the data
and the model also provides a self-consistent framework within which
the other of the observational properties of this group of objects can
be understood. Motivated by the need to balance the energy budget of
the outer, line-emitting part of the disk, Chen \& Halpern (1989)
proposed that the inner accretion disk is a vertically extended
structure that can illuminate the outer disk effectively and power the
line emission. The energy budget problem comes about because the H\a\
luminosity is uncomfortably close to, or in some cases in excess of,
the power dissipated by viscous stresses in the line-emitting portion
of the disk. This power deficit is demonstrated in Table~\ref{Tdisk}
(an updated version of Table~9 of paper~I) and in Figure~\ref{Fdisk},
where we compare their H\a\ luminosities with the power output of
their line-emitting disks. The power output of the disk was computed
using equation (4) of paper~I and the X-ray luminosity, under the
assumption that the X-ray luminosity represents 1\% of the available
accretion power.  In 30\% of the cases the H\a\ luminosity is higher
than the local power output of the line-emitting portion of the disk
and in 90\% of the cases this fraction is more than 0.1
(Figure~\ref{Fdisk}a). On the other hand, the soft X-ray luminosity
always exceeds the H\a\ luminosity (their ratio is more than 10 in
70\% of the cases). This comparison suggests very strongly that the
Balmer lines are powered by photoionization of the outer disk by
radiation produced in the inner disk.

The inner accretion disk was hypothesized by Chen \& Halpern (1989) to
be a hot, ion supported torus (Rees et al. 1982) which radiates the
available accretion power inefficiently, and its existence is favored
at low accretion rates (one to two orders of magnitude below the
Eddington rate). The preferential association of double-peaked
emitters with radio-loud AGNs was also explained automatically, since
the original ion torus model was intended to explain the formation of
radio jets. The inner radius of the line-emitting part of the disk
was, therefore, identified with the radius at which the disk puffs up
to form the torus, while the outer radius was identified with the
radius at which the disk fragments into discrete, self-gravitating
clouds. The structure and properties of the ion torus are very similar
to those of advection-dominated or convection-dominated accretion
flows (ADAFs and CDAFs, possibly accompanied by an outflow) discussed
more recently by many authors, including Narayan \& Yi (1994, 1995),
Blandford \& Begelman (1999), Igumenchev \& Abramowicz (1999, 2000),
and Narayan, Igumenchev, \& Abramowicz (2000) \footnote{We will use
the term ``ion torus'' throughout to denote a generic hot, vertically
extended, optically thin, and radiatively inefficient accretion flow.
These general properties, rather than the exact structure of the flow
are what is important to our interpretation.}.

The SED of an ion torus is relatively hard, lacking a ``UV bump'' (the
hallmark of a geometrically thin accretion disk), having instead a
shape that resembles a power law in the UV and soft X-ray bands, and a
peak in the far-IR, between 30 and 300~$\mu$m (see, for example, Di
Matteo et al. 1999, 2000, and Ball, Narayan, \& Quataert 2001). Such
an SED explains the weakness of the non-stellar continuum in the
optical spectra of double-peaked emitters in general. Moreover, the IR
peak in the SEDs of Arp~102B and 3C~390.3 can be identified with the
peak predicted by emission models for an ion torus. The unusually low
ionization state of the narrow-line regions of double-peaked emitters
(see \S\ref{S_optical}) can also be understood in this context and
with the help of photoionization models for LINERs (Halpern \& Steiner
1983; Ferland \& Netzer 1983). These models show that an otherwise
``normal'' narrow-line region, illuminated by an ionizing continuum
with a power-law spectrum and a low ionization parameter produces
emission lines with relative strengths very similar to what is
observed in double-peaked emitters.  More recent photoionization
calculations, carried out by Nagao et al. (2002) and intended
specifically for double-peaked emitters, reach the same conclusion.
In particular, Nagao et al. (2002) have compared the narrow-line
ratios of double-peaked emitters and of a large collection of other
radio-loud AGNs with photoionization models employing two different
SEDs: one consisting of a simple power law and one including a ``UV
bump.'' They find that the narrow-line ratios of double-peaked
emitters are best explained by models with a power-law SED.

\subsection{Further Tests of the Accretion Disk 
Interpretation\label{S_disc_tests}}

The origin of the broad, double-peaked Balmer lines in an accretion
disk is supported further by tests on two of the best studied objects,
Arp~102B and 3C~390.3, which we describe below.

In the case of Arp~102B, the profile of the Ly$\alpha$ line is
single-peaked, bell-shaped, and considerably narrower than those of
the double-peaked Balmer lines (see Figure~3 of Halpern et
al. 1996). The best interpretation of this difference is that the
region producing the double-peaked Balmer lines has such a high
density and column density ($n \sim 10^{15}~{\rm cm}^{-3}$ and $N_{\rm
H} \sim 10^{25}~{\rm cm}^{-2}$) that the Ly$\alpha$ photons it
produces are trapped by resonance scattering and destroyed by
collisional de-excitation. In fact, this dramatic difference in line
profiles was {\it predicted} by Rokaki, Boisson, \& Collin-Souffrin
(1992) based on the accretion disk photoionization models of
Collin-Souffrin \& Dumont (1989).

A reverberation mapping campaign of 3C~390.3 by the {\it International
AGN Watch} (Dietrich et al. 1998; O'Brien et al. 1998) has shown that
the Balmer lines respond to variations of the continuum with a lag of
20 days and that the blue and red sides of the line respond together
with an upper limit on the delay between them of 4~days. Similar
results were obtained by Sergeev et al. (2002), who also estimated the
transfer function of the H\a\ line of 3C~390.3 and found it to peak
sharply at a lag of 60 days.  This is exactly the behavior that one
would expect from an accretion disk, as noted by Gaskell (1999).
Moreover, Sergeev, Pronik, \& Sergeeva (2000) find the same behavior
in Arp~102B, with the broad H\a\ line lagging the continuum variations
by less than 66 days. Estimates of the mass of the central black hole
in 3C~390.3 from reverberation mapping range from $3\times 10^{8}~{\rm
M}_{\odot}$ (Peterson \& Wandel 2000) to $2\times 10^{9}~{\rm
M}_{\odot}$ (Sergeev et al. 2002).

Finally, we note a reassuring consistency check based on the
superluminal motion observed in the radio core of 3C~390.3 by Alef et
al. (1994, 1996). The combination of a number of constraints on the
orientation of the radio jet yields an inclination angle of the jet of
19$^{\circ}$--42$^{\circ}$ (Eracleous, Halpern, \& Livio 1996
\footnote{Eracleous et al. (1996) used the observed superluminal speed
to derive an upper limit on the jet inclination angle of 33\deg, under
the assumption that $H_0=50$~km~s\m1~Mpc\m1. If we take
$H_0=70$~km~s\m1~Mpc\m1, we obtain an upper limit of 42\deg\ on the
jet inclination angle.}).  This is in good agreement with the
inclination angle of the axis of the accretion disk of
$26_{-2}^{+4}$~degrees obtained in paper~I by fitting the H\a\
profile.

The observations described in this section also serve as tests of 
alternative scenarios for the origin of double-peaked lines, which 
we discuss in later sections.

\subsection{Discussion of Individual Objects: Pictor~A and 
CSO~643\label{S_indiv}}

In \S\ref{S_fit_Ha} we noted that the simplest, axisymmetric disk
model does not describe the H\a\ profiles of Pictor~A and CSO~643
extremely well, which led us to use more sophisticated models. Here we
describe the physical motivation for the models that we used.

\noindent
{\it Pictor A}. -- The double-peaked lines appeared some time
between 1983 and 1989. A spectrum taken in 1983 by Filippenko (1985)
shows a single-peaked H\a\ line with very broad wings, while spectra
taken between 1989 and 1994 (Halpern \& Eracleous 1994; Sulentic et
al. 1995a) show the H\a\ line to be double-peaked and quite
variable. Halpern \& Eracleous (1994) speculated that the appearance
of the double-peaked line was related to a major structural change in
the accretion disk, while Sulentic et al. (1995a) attributed the
double-peaked lines to a bipolar outflow. An important feature of the
H\a\ profile of Pictor~A is that the red peak is stronger than the
blue (at least at the time that our spectrum was taken), contrary to
the prediction of the simplest disk models. This feature, however,
does not preclude the origin of the line in an accretion disk.  Any
departure of the disk from axisymmetry will lead to line profiles that
differ from those predicted by the simplest models. Specific examples
of non-axisymmetric disks include (but are not limited to) disks with
spiral waves (Chakrabarti \& Wiita 1994), eccentric disks (Eracleous
et al. 1995), and disks with warps induced by intense irradiation
(Pringle 1996; Maloney, Begelman \& Pringle 1996). In fact, the
dynamical signatures of the first two of these examples of
non-axisymmetric disks have been detected in cataclysmic variables
(Steeghs, Harlaftis, \& Horne 1997; Baptista \& Catal\'an 2000;
Patterson, Halpern, \& Shambrook 1993).  Refining the simplest disk
model by introducing an eccentricity allows us to fit the H\a\ profile
of Pictor~A, as illustrated in Figure~\ref{Ffit}c. Such departures
form the simplest model are not uncommon, as shown by Eracleous et
al. (1995). We must also note that the H\a\ profile of Pictor~A has
been varying dramatically over the past decade (see the sequence of
spectra presented by Eracleous \& Halpern 1998), which means that the
particular models that we discuss here are at best applicable to a
single epoch.

\noindent {\it CSO 643}. -- The double-peaked H\a\ profile of CSO~643
resembles those of 3C~59, PKS~0340--37, and CBS~74: they all have a
pronounced red shoulder, which is much more extended than the blue
one, and require small inclination angles ($i<20^{\circ}$) for a good
fit. However, in the CSO~643 the net redshift of the entire profile is
smaller than in the other objects preventing us from obtaining a good
fit with a simple circular-disk model. There are two ways to improve
the fit:
\hfill\break (a)
We can apply an {\it ad hoc} blueshift to the model of 1370~\kms. The
best-fitting circular disk model found under this assumption is 
shown in Figure~\ref{Ffit}e overlayed on the observed profile. The
model parameters are summarized in Table~\ref{Tfit}.
\hfill\break (b)
We can use a more sophisticated model with more free parameters, such
as an elliptical disk model (Eracleous et al. 1995). The best-fitting
model of this type is shown in Figure~\ref{Ffit}g and the model
parameters are summarized in Table~\ref{Tfit}.
\hfill\break 
Either of the above modifications to the simple disk model
can be physically justified by appealing to the effects of an unseen
supermassive companion to the accreting black hole (see, for example,
Rees et al. 1982). In case (a) the blueshift is a result of the
orbital motion of the accreting black hole around the center of mass
of the supermassive binary. In case (b) the eccentricity of the
hypothesized elliptical disk is excited by the tidal effect of the 
unseen companion (see the discussion in Eracleous et al. 1995 and 
references therein).

\subsection{Alternative But Less Appealing Scenarios for 
Double-Peaked Lines\label{S_disc_alt}}

In addition to accretion-disk emission three other scenarios have been
suggested for the origin of double-peaked emission lines, namely (a)
emission from a binary broad-line region associated with a
supermassive binary black hole, (b) emission from the oppositely
directed sides of a bipolar outflow, and (c) emission from a
spherically symmetric broad-line region illuminated by an anisotropic
source of ionizing radiation. In this section we describe these
scenarios and evaluate them in the light of the currently available
data. We consider all of the observational consequences of each
scenario and we assess it based on how well it can explain the
observational properties of double-peaked emitters.

\subsubsection{Binary Broad-Line Region} 

The idea that supermassive binary black holes (resulting from the
merger of their parent galaxies) reside in the nuclei of radio-loud
AGNs was suggested by Begelman, Blandford, \& Rees (1980), who also
noted that double-peaked emission lines could be an observational
consequence of such a scenario. Gaskell (1983, 1988, 1996a,b)
identified a number of radio-loud AGNs with displaced broad emission
lines as candidate supermassive binaries and proposed that
supermassive binaries are very common in both radio-loud and
radio-quiet AGNs with double-peaked emitters being the most obvious
cases. In this scenario the spectroscopic properties of the
double-peaked emitters should not differ from the those of the average
(radio-loud) AGN, which is in contradiction with the observational
properties described in \S5 and \S\ref{S_disc_tests}
(most notably with the dramatic difference between the profiles of the
H\a\ and Ly\a\ lines of Arp~102B and the reverberation of the H\a\
line of 3C~390.3).

Another serious disagreement between this scenario and the
observations results from its prediction that the two peaks of a
double-peaked line should drift in opposite directions as a result of
the orbital motion of the binary.  Although the expected orbital
periods can range from decades to centuries, the signature of orbital
motion can be detected, in spectra spanning one or two decades. Thus
Halpern \& Filippenko (1988, 1992) searched for radial velocity
variations in the displaced peaks of Arp~102B and 3C~332, using
spectra spanning a decade, but did not find any. A more extensive
study of Arp~102B, 3C~390.3, and 3C~332 by Eracleous et al. (1997) did
not find any evidence for orbital motion; a suggestive variability
trend found by Gaskell (1996a) in 3C~390.3 lasted only until 1988
(Eracleous et al. 1997; Shapovalova et al. 2001; Sergeev et al. 2002)
The lack of evidence for orbital motion yielded lower limits on the
binary masses of \ten{10}--\ten{11}~\Msol, which are in conflict with
a number of other observations and with theory, leading to the
rejection of this scenario as a general explanation for double-peaked
emission lines (see Eracleous et al. 1997 for a more detailed
discussion).

\subsubsection{Emission from a Bipolar Outflow} 

Since double-peaked emitters are radio-loud AGN it is conceivable that
the double-peaked lines arise in a bipolar outflow resulting from the
interaction of the powerful radio jets with gas immediately around the
central engine. Norman \& Miley (1984) discussed the interaction of
the jets with the emission-line regions in qualitative terms, while
Zheng, Binette, \& Sulentic (1990) constructed quantitative models for
the profiles of emission lines produced in the outflow. Such models
were applied to the double-peaked H\a\ lines of 3C~390.3 and Pictor~A
by Zheng, Veilleux \& Grandi (1991) and by Sulentic et al. (1995a),
respectively.  Bipolar outflow models are not immediately appealing,
however, because statistical studies of the H\b\ profiles of
radio-loud AGNs suggest that the line-emitting gas is arranged in a
flat disk perpendicular to the radio axis (see, for example, Wills \&
Browne 1986; Jackson \& Browne 1991b; Brotherton 1996; Marziani et al.
1996).  The conclusions of these studies and the fact that most
double-peaked emitters are associated with double-lobed radio sources
argue against bipolar outflows and contradict the suggestion by
Sulentic et al. (1995a) that double-peaked emitters represent an
extreme segment of the radio-loud AGN population in which the axis of
the outflow is oriented very close to the line of sight.  The
following additional observational constraints make bipolar outflow
models for double-peaked emission lines unlikely.

\begin{enumerate}

\item
The dramatic difference between the observed Ly\a\ and H\a\ profiles
of Arp~102B (Halpern et al 1996) cannot be easily explained. The
radial velocity gradient in the outflow should reduce the optical
depth of Ly\a\ photons and allow them to escape easily with the
consequence that Ly\a\ and the Balmer lines should have similar
profiles.

\item
Reverberation mapping of 3C~390.3 (see \S\ref{S_disc_tests}) has shown
that both sides of the H\a\ line respond together to changes in the
ionizing continuum, contrary to the expectation that the blue-shifted
side of the line should respond first, with hardly any lag from the
continuum variations.

\item
If the double-peaked lines of 3C~390.3 originate in a bipolar outflow,
the reverberation results reviewed in \S\ref{S_disc_tests} imply that
the outflow should be viewed nearly at right angles to its axis
($i>84^{\circ}$, following Livio \& Pringle 1996 and Livio \& Xu
1997). However, this orientation is in contradiction with the observed
radio properties of 3C~390.3 (see discussion in \S\ref{S_disc_tests}).

\end{enumerate}

\subsubsection
{Spherically Symmetric Broad-Line Region, Illuminated Anisotropically}

This model was discussed in its general form by Goad \& Wanders (1996)
although it was originally considered by Wanders et al. (1995) as an
interpretation of the reverberation mapping results for the Seyfert
galaxy NGC~5548. The broad-line region is assumed to consist of a
large number of clouds in randomly inclined Keplerian orbits occupying
a thick spherical shell. The clouds are photoionized by a central
source (presumably an accretion disk) which emits anisotropically. In
the specific picture discussed by Goad \& Wanders (1996) the
anisotropic emission was described by two conical beams which could
possibly be superposed on an isotropic ``background''
illumination. For specific combinations of the beam opening angle and
orientation of the observer relative to the axis of the beam, the
resulting emission-line profiles appear double-peaked and the expected
frequency of double-peaked lines is consistent with what is observed.
It was also suggested that the UV spectrum of Arp~102B could be
explained in this context if the ionizing continuum consists of a hard
X-ray power-law emitted isotropically and a softer, ``UV bump''
spectrum making up the two conical beams (cf, Netzer 1987). Moreover,
high- and low-ionization lines would, in general, have different
profiles.  Thus, an observer whose line of sight is within the cone of
the UV beam would find the the Ly\a\ and \ion{C}{4}\ lines to be
relatively narrow and single peaked, but the Balmer and \ion{Mg}{2}\
lines would appear very broad and double peaked, just as observed in
Arp~102B. The weakness of the non-stellar continuum in the optical
spectra of double-peaked emitters could also be explained by placing
the observer's line of sight outside of the cones of the UV beams.

Although this scenario has its attractions, its applicability to
double-peaked emitters does not withstand close scrutiny.  It is at 
odds with the observations in the following ways:

\begin{enumerate}

\item
It does not provide a self-consistent explanation of all of the properties of
Arp~102B: in order to explain the difference between the profiles of
the Ly\a\ and Balmer, the observer's viewing direction must lie within
the conical UV beams, but to explain the weakness of the non-stellar
continuum relative to the starlight, the observer's viewing direction
must be outside the conical UV beams.

\item
It does not explain the unusually strong low-ionization narrow
lines of double-peaked emitters.

\end{enumerate}

From a theoretical perspective the physical basis for the assumed
cloud distribution and kinematics is questionable. Even if such a
system of clouds on randomly inclined, Keplerian orbits could be
produced at all, it would be destroyed by collisions in 100 dynamical
times, or in 200 years in the case of 3C~390.3 (see the general
estimate and discussion in Matthews \& Capriotti 1985).  An
additional, and arguably more catastrophic, physical effect is the
decay of the orbits of discrete clouds due to a Poynting-Robertson
drag, which could occur on a time scale of 6--60 years (see Mathews \&
Capriotti 1985). We therefore doubt that this picture of the
broad-line region is physically realizable.

\subsection{Connection of Double-Peaked Emitters With the Greater AGN 
Population.\label{S_disc_single}}

If the H\a\ emission line profiles of approximately 20\% of radio-loud
AGNs are to be attributed to accretion disks, what of the remaining
objects?  Our understanding of the dynamics of the line-emitting gas
would improve tremendously if a single model could be found that would
describe the majority of emission-line profiles, especially if the
same model can explain double-peaked profiles.  As such, double-peaked
emission lines place important constraints on any model that
seeks to explain AGN broad-line regions in a universal way.

A possible new clue is the discovery of
double-peaked Balmer emission lines in many LINERs, which underscores
the connection between the two types of hosts. Examples of LINERs with
double-peaked emission lines include NGC~1097 (Storchi-Bergmann et al.
1993), M81 (Bower et al. 1997), NGC~4203 (Shields et al. 2000),
NGC~4450 (Ho et al. 2000), and NGC~4579 (Barth et al. 2001). The
prototypical double-peaked emitter, Arp~102B is also a LINER
(Stauffer, Schild, \& Keel 1983).  Since LINERs are found in about
30\% of all nearby galaxies and a significant fraction of them could
be AGNs, the incidence of double-peaked emission lines among the
entire AGN population could be quite high. The exact fraction of AGNs
among LINERs and the frequency of double-peaked emission lines in
LINERs remain to be quantified, however.

Motivated by these considerations, we explore here whether accretion
disk models, which offer the best explanation for double-peaked lines,
can also explain the single-peaked profiles observed in the majority
of AGNs. In support of this approach, we note that our comparison of
the widths and shifts of double-peaked and single-peaked emission
lines suggests a similarity between the corresponding broad-line
regions.  Furthermore, if we attribute the mean redshift measured over
all radio-loud AGNs to the combined effects of gravity and transverse
motions of the line-emitting gas in a Keplerian disk, it allows us to
infer a characteristic distance of the line production site from the
central object of $6,000\; r_{\rm g}$, comparable to the outer radii
of the line-emitting disks found in \S4 and in paper~I. This
estimate supports the above suggestion that all radio-loud AGNs may
harbor line-emitting accretion disks and/or additional line-emitting gas
in related structures (e.g., an accretion-disk wind).

The following possibilities for the origin of single-peaked lines in
accretion disks have been discussed in the literature:

\begin{enumerate}

\item
The line-emitting disk in most AGNs could be quite large with a ratio
of outer to inner radius of order 10 or more. This would bring the two
peaks of a double-peaked line close together and make the profiles
appear single-peaked (Dumont \& Collin-Souffrin 1990a; Jackson,
Penston, \& P\'erez 1991). To illustrate this point we have simulated
H\a\ spectra by assuming that the broad H\a\ line is produced in a
large accretion disk, namely one with an inner radius of order a few
hundred $r_{\rm g}$ and an outer radius greater than $10^4\; r_{\rm
g}$ i.e., $\xi_2/\xi_1\sim 10-100$.  To complete the simulation we
added the usual narrow lines in the vicinity of H\a\ and Poisson
noise. The narrow lines were assumed to have Gaussian profiles with a
FWHM of a few hundred \kms, while the noise was generated assuming a
$S/N$ of 30--200 in the continuum. We explored a range of inclination
angles between 10\deg\ and 40\deg\ and a range of emissivity power-law
indices between 1.5 and 3. In Figure~\ref{Fsimul} we show a montage of
simulated spectra spanning a range of parameter values.  These
examples were chosen because they resemble very closely some of the
observed H\a\ spectra presented in paper~I.  Thus, large accretion
disk models can reproduce, at least qualitatively, a wide variety of
profiles observed in radio-loud AGNs.  Moreover, red asymmetries are
quite common, particularly below half maximum of the profile,
which is a desirable
feature since this is commonly observed. The large outer radius of the
line emitting disk and the H\a+[\ion{N}{2}] narrow-line complex can
conspire to hide the characteristic twin shoulders that are associated
with lines from a disk. We find that ``shallow'' emissivity profiles
(described by power-law indices $q\approx 1.5 - 2$) produce line
profiles that bear a closer resemblance to the observed ones than
those produced by ``steep'' ($q\approx 3$) emissivity laws.  In the
context of the photoionization calculations of Collin-Souffrin \&
Dumont (1989) this can be understood as the result of a very luminous
central source.

\item
The disks in most AGNs could be oriented close to face on. This
possibility was explored by Corbin (1997a) who computed the profiles
resulting from nearly face-on, flattened broad-line regions resembling
face-on disks. Some of the simulated H\a\ profiles that we present in
Figure~\ref{Fsimul} also correspond to face-on disks. The main
conclusion from these calculations is that face-on disk models
reproduce many of the desired properties of observed line profiles,
including a single peak, extended red wings (i.e., redward
asymmetries), and a net redshift of the entire line.

\item
Double-peaked line profiles from a disk can easily be turned into
single-peaked profiles by the presence of a disk wind. Murray \&
Chiang (1997) have shown that although the outflow velocity of the
wind is much smaller than the rotational velocity, its velocity
gradient is as large as the rotational velocity gradient. The main
physical consequence is that photons can escape much more easily along
lines of sight with a small projected velocity. The resulting line
profiles are single peaked with broad wings even though the emission
comes from gas that is essentially on circular orbits. This was
illustrated clearly by Murray \& Chiang (1997) who computed profiles
of UV resonance lines arising in a wind and compared them to observed
line profiles of quasars. An additional illustration was provided by
Chiang \& Murray (1996) who showed that the observations of
reverberation in the \ion{C}{4} line of NGC~5548 could also be
explained, at least qualitatively, in the context of an accretion-disk
wind model.

\end{enumerate}

Of the possibilities discussed above, we favor the accretion-disk wind
scenario because of a number of additional appealing features that it
possesses, as we discuss further below. In particular, it offers a way
of connecting double-peaked emitters to the greater AGN population.
It also provides an explanation for the H\a\ blueshifts observed in
some double-peaked emitters (see Figure~\ref{Fshifts} and \S3), since
a low-optical depth wind can impart a blueshift on lines from the disk
without altering their double-peaked profiles. We therefore propose
that the wind is the broad-line region in most AGNs, which accrete at
rates that are a sizeable fraction of the Eddington rate. This idea is
by no means new; it was suggested, for example, by Murray et
al. (1995) to explain broad-absorption line quasars and by Elvis
(2000) to explain the absorption features observed in the X-ray
spectra of quasars. The dynamics of the wind were worked out
analytically by Murray et al. (1995) and through detailed numerical
simulations by Proga, Stone, \& Drew (1999) and Proga, Stone, \&
Kallman (2000), which confirm the general analytic results.

To show how double-peaked emitters and low-luminosity AGNs fit into
this scheme, we note that the structure of the accretion flow is
controlled largely by the Eddington ratio, i.e., the ratio of the
accretion rate to the Eddington rate ($\dot M / \dot M_{\rm Edd}$). At
large Eddington ratios ($\dot M / \dot M_{\rm Edd}\gs 0.1$) the inner
accretion disk is geometrically thin and optically thick, as described
by the model of Shakura \& Sunyaev (1973). Its emitted SED has a
prominent ``UV bump,'' which can exert a substantial radiation
pressure on the atmosphere of the outer disk via resonance line
absorption, accelerating a wind. We posit that such a structure
applies to Seyfert galaxies and quasars. At the opposite extreme of a
low Eddington ratio ($\dot M / \dot M_{\rm Edd}\ls 10^{-2}$) the inner
disk changes to an ion torus whose emitted SED is a hard X-ray
power-law without a ``UV bump''. When such a hard spectrum impinges on
the atmosphere of the outer disk it ionizes it and reduces the
potential of exerting radiation pressure via resonance line absorption
(see, for example, the discussion by Murray et al. 1995). The
potential for a significant radiation pressure is reduced further by
the lack of substantial numbers of UV photons in the incident
SED. Thus the wind becomes feeble and the accretion disk is unveiled,
allowing the observer to see double-peaked emission lines from the
disk proper. In such objects the feeble wind would be the primary
source of UV resonance emission lines, which would be single peaked as
a result of their origin (see the discussion by Collin-Souffrin \&
Dumont 1989). Moreover, the wind may also be observable through
absorption lines that it produces. Such wind-like absorption lines
have actually been detected in Arp~102B (Halpern et al. 1996;
Eracleous, Halpern, \& Charlton 2003) and other double-peaked emitters
(Eracleous 2002).

\section{Epilogue\label{S_epilog}}

We have presented the completion of our survey of moderate redshift,
radio-loud AGNs whose primary motivation was to search for more
examples of double-peaked emission lines.  We find that 20\% of the
objects surveyed have H\a\ lines with double peaks or twin
shoulders. In 17 of the 24 cases the blue peak/shoulder is stronger
than the red.  The H\a\ profiles of 13 of these 17 objects can be
fitted quite well by the simplest possible disk model (homogeneous and
axisymmetric). The H\a\ profiles of many of the remaining objects can
be fitted by more sophisticated disk models (non-axisymmetric and/or
inhomogeneous). Double-peaked emitters possess a number of additional
properties which distinguish them from the average radio-loud AGN and
make them similar to accretion-powered LINERs. A consideration of all
of the available data shows an origin of double-peaked emission lines
in an accretion disk to be the preferred interpretation. Alternative
scenarios are rather unsatisfying: although they can produce
double-peaked emission lines, at least in principle, they cannot
explain all of the properties of double-peaked emitters, they fail
some direct observational tests, and in some cases their physical
foundations do not withstand close scrutiny.

A physical model in which the inner accretion disk is an ion torus
that illuminates the outer disk can explain the observed properties of
double-peaked emitters. This model also provides a framework in which
the connection between double-peaked emitters and the greater AGN
population can be sought. More specifically, we propose that in most
AGNs the broad-line region is an accretion-disk wind, while in
double-peaked emitters and other low-luminosity AGNs (such as some
LINERs) the broad-line region is the outer accretion disk. This
transformation is brought about by a decline of the accretion rate
from a sizable fraction of the Eddington rate to values that are more
than two orders of magnitude lower. At these extremely low accretion
rates the structure of the inner disk changes to an ion torus and the
wind that cloaks the outer disk diminishes. This scenario can be
tested by comparing the profiles of UV resonance lines with the
profiles of the Balmer lines of double-peaked emitters.

Even though our survey targeted radio-loud AGNs so as to maximize the
efficeincy of finding double-peaked emitters, radio-quiet AGNs also
host double-peaked emission lines. As we noted in
\S\ref{S_disc_single} a significant number of LINERs have recently
been found to possess double-peaked Balmer lines. LINERs have been
traditionally regarded as radio-quiet objects but it was recently
argued by Ho et al. (2000) and Terashima \& Wilson (2003) that the
ratio of the radio luminosity of LINERs to their non-stellar optical
luminosity is considerably higher than that of radio-quiet AGNs.  The
much larger sample of 116 double-peaked emitters found in the Sloan
Digital Sky Survey (SDSS; see Strateva et al. 2003) comprises 76\%
radio-quiet AGNs, indicating that radio-quiet double-peaked emitters
are more numerous than radio-loud ones. However, double-peaked
emitters are 1.6 times more likely to be found among radio-loud AGNs.

Finally, although it is beyond the scope of this paper, variability of
double-peaked profiles deserves discussion.  We distinguish two types
of variability (a) short-term variability of the line flux due to
reverberation of the ionizing continuum (time scales of order weeks),
and (b) long term variability of the line profiles (time scales of
order months to years). In this paper we have made extensive use of
the results of reverberation mapping of 3C~390.3 (Dietrich et
al. 1998; O'Brien et al. 1998; Sergeev et al. 2002), because their
implications are straightforward and important.  Since such
experiments have been performed only for 3C~390.3, we have taken these
results to be representative of all double-peaked emitters. However,
reverberation mapping of additional objects is sorely needed to check
their behavior and to measure the dimensions of the line-emitting
region.

Studies of long-term profile variations of the line profiles have been
carried out for a handful of double-peaked emitters, namely 3C 390.3,
(e.g., Zheng et al. 1991; Eracleous et al. 1997; Gilbert et al. 1999;
Sergeev et al. 2002) Arp 102B (e.g., Miller \& Peterson 1990; Newman
et al. 1997), NGC 1097 (Storchi-Bergmann et al. 1995, 1997, 2003),
3C~332 (Eracleous et al. 1997; Gilbert et al. 1999), and profile
variability results on Pictor~A have also been reported (Halpern \&
Eracleous 1994; Sulentic et al. 1995a; Eracleous \& Halpern 1998) but
no comprehensive study exists yet.  The long-term variability patterns
show a wide variety, although the most pronounced and obvious pattern
is the slow and systematic modulation of the relative strengths of the
two peaks. In this paper we have appealed to profile variations
primarily to test the binary black hole hypothesis. However, the
interpretation of the long-term profile variations is not yet clear
and they do not yet constitute a straightforward test of scenarios for
the origin of the lines. Such slow variations are certainly not the result
of reverberation of a variable ionizing continuum since the
light-crossing time of the line-emitting region is of order a few
weeks (see \S\ref{S_disc_tests}).  More generally, reverberation
mapping of Seyfert galaxies has shown that although the {\it
integrated fluxes} of broad emission lines do vary in response to a
variable ionizing continuum, the line profiles do not (e.g., Ulrich et
al. 1991; Wanders \& Peterson 1996; Kassebaum et al. 1997). The time
scales on which the {\it line profiles} change significantly are
longer than the dynamical time of the line-emitting region by a factor
of several. Now that accretion disk emission is emerging as the
favorite scenario for the origin of double-peaked emission lines, we
will be able to investigate the origin of the observed profile
variations through detailed case studies and exploit them to learn
about dynamical phenomena in AGN accretion disks.

\acknowledgements 

We thank the anonymous referee for helpful comments and suggestions.
We are grateful to M. S. Brotherton and B. J. Wills for allowing us to
use their measurements of \ion{Mg}{2}/H\b\ ratios of many radio-loud
AGNs prior to publication. We also thank A. V. Filippenko for
obtaining some of the spectra presented in this paper at Lick
Observatory and we acknowledge the help of A.  Barth in the reduction
of these spectra. During the early stages of this work M.E. was based
at the University of California, berkeley, and was supported by Hubble
Fellowship grant HF-01068.01-94A awarded by the Space Telescope
Science Institute, which is operated by the Association of
Universities for Research in Astronomy, Inc., for NASA under contract
NAS~5-26555.


\clearpage

{\rotate
\begin{figure}
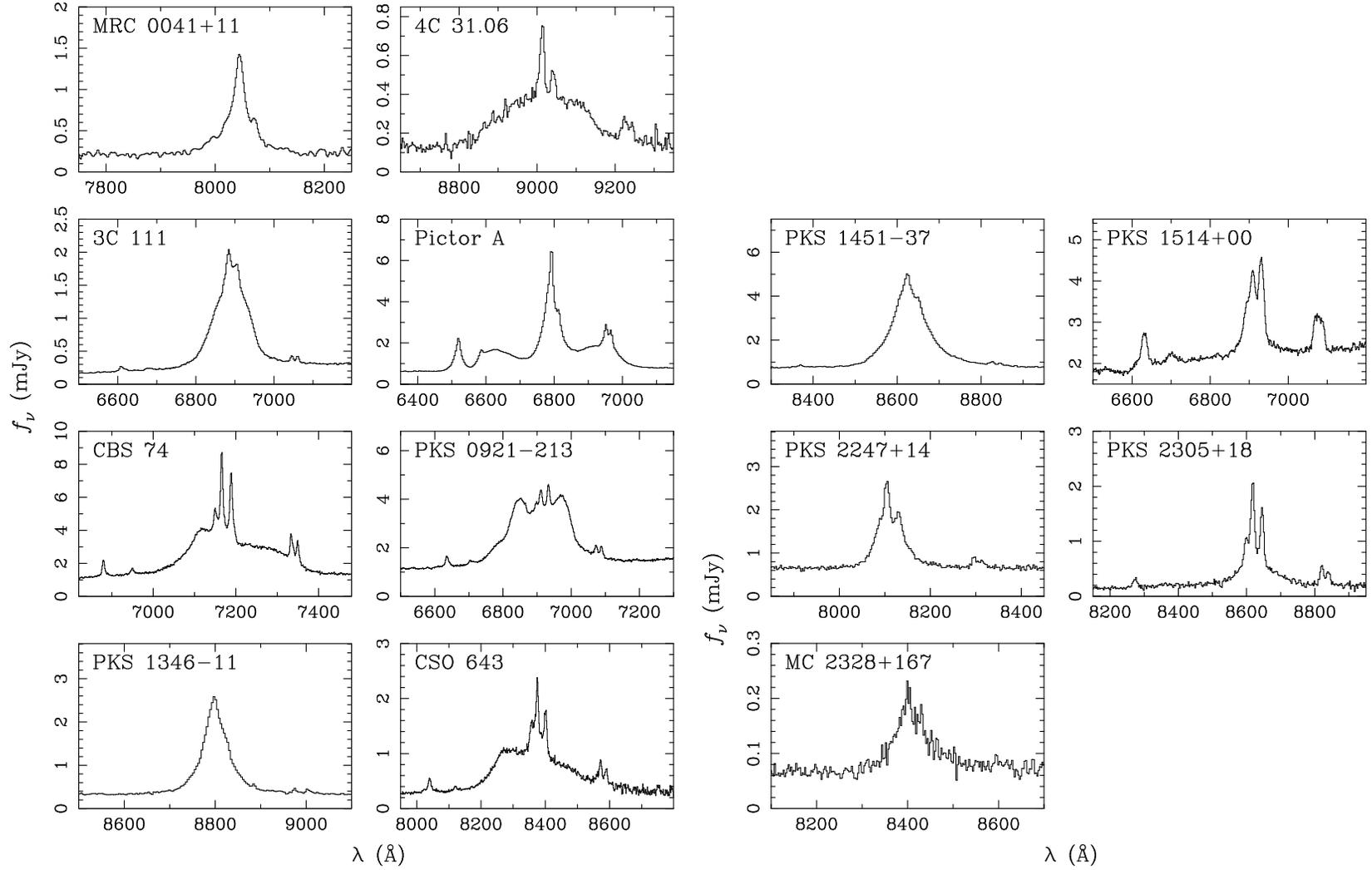
 
\epsscale{1.2}
\begin{minipage}[t]{3.5in}
\hskip -0.5truein
\leftline{\plotone{f1a.eps}}
\end{minipage}
\hskip 0.3truein
\begin{minipage}[t]{3.5in}
\leftline{\plotone{f1b.eps}}
\end{minipage}
\caption{ The H\a\ spectra of the 13 newly-observed broad-lined
objects. A list of these objects and the journal of observations are
given in Table~\ref{Tjour}.
\label{Fspec}}
\end{figure}
\begin{figure}    
\epsscale{1.4}
\begin{minipage}[t]{3.5in}
\hskip -1truein
\leftline{\plotone{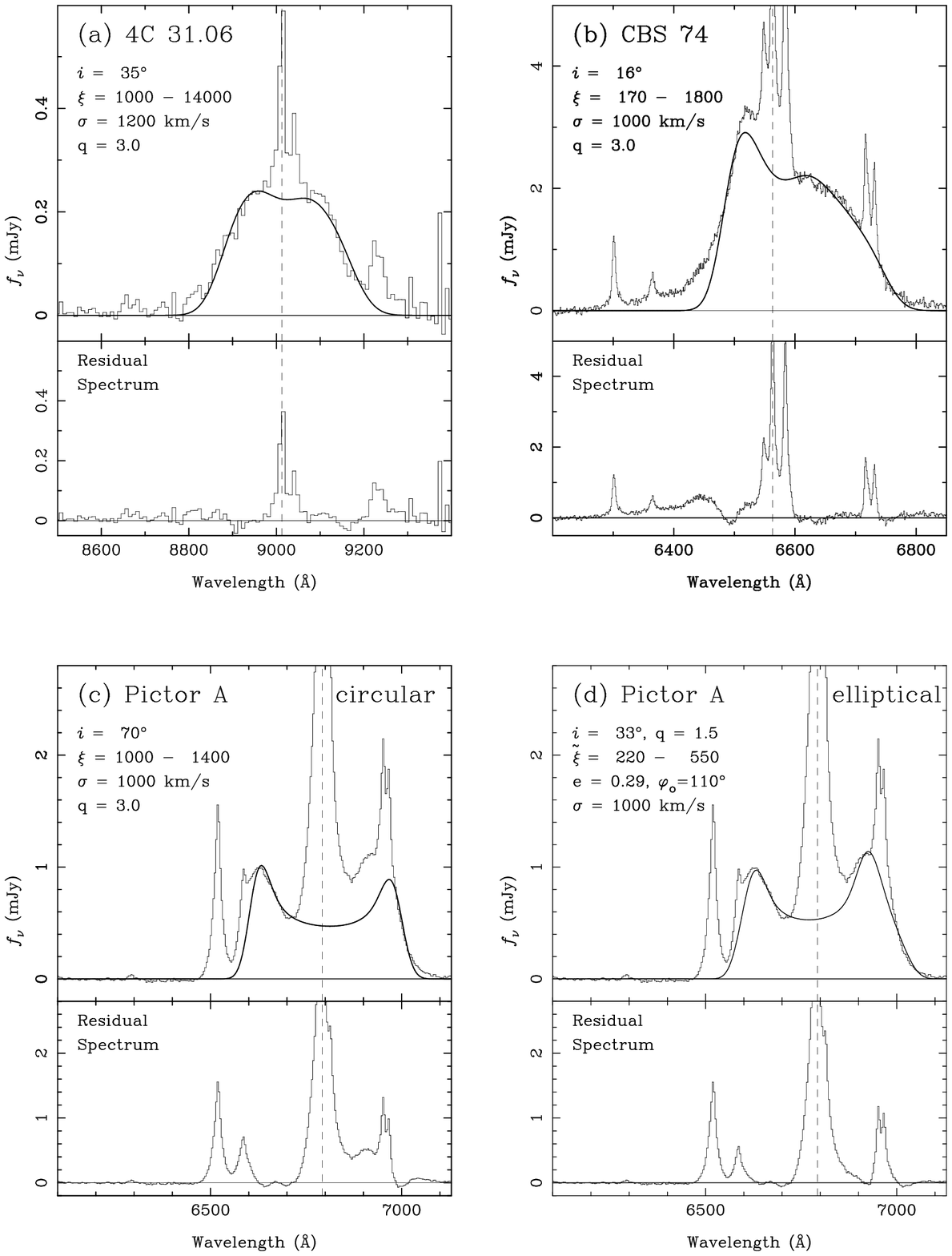}}
\end{minipage}
\begin{minipage}[t]{3.5in}
\leftline{\plotone{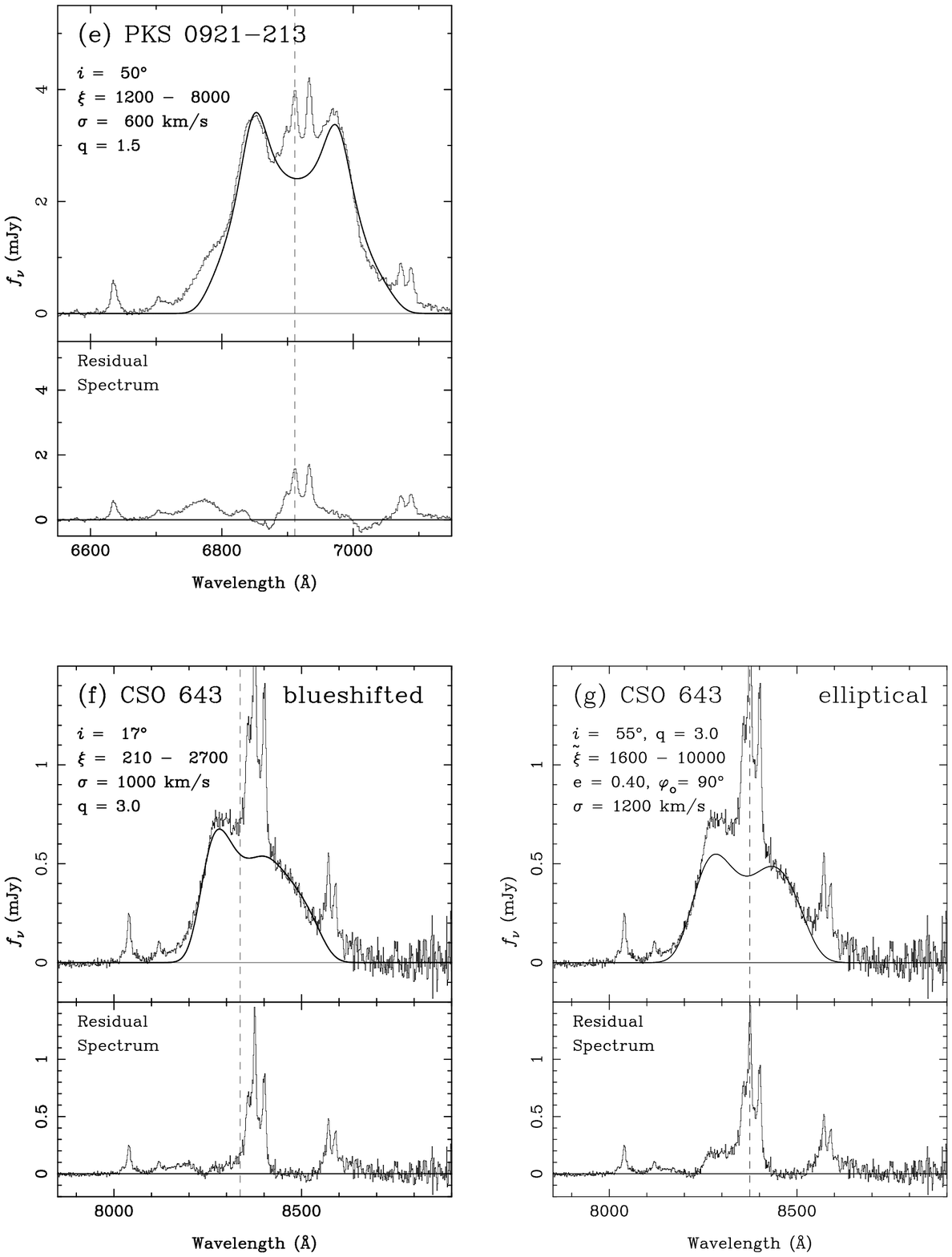}}
\end{minipage}
\caption{Fits of relativistic, Keplerian disk models to objects with
double-peaked H\a\ lines. The model parameters are discussed in
\S\ref{S_fit_Ha} of the text and their best fitting values are given
in each figure and summarized in Table~\ref{Tfit}.  The objects are
arranged in order of increasing right ascension.  In every case the
upper panel shows the H\a\ spectrum of the object, after subtraction
of the continuum, with the best fitting model superposed (thick
solid line). The lower panel shows the residual after subtraction of
the model. The dashed vertical line marks the wavelength adopted 
for the model profile. In panels (d) and (g) we show a fit of an eccentric
disk model to the H\a\ profile of Pictor~A and CSO~643, while in panel
(f) we show a fit of a blueshifted circular disk model to the profile
of CSO~643 (see \S\ref{S_fit_Ha} for a detailed discussion).
\label{Ffit}}
\end{figure}
\begin{figure}   
\hskip -1truein
\begin{minipage}[t]{4.in}
\epsscale{1.3}
\centerline{\plotone{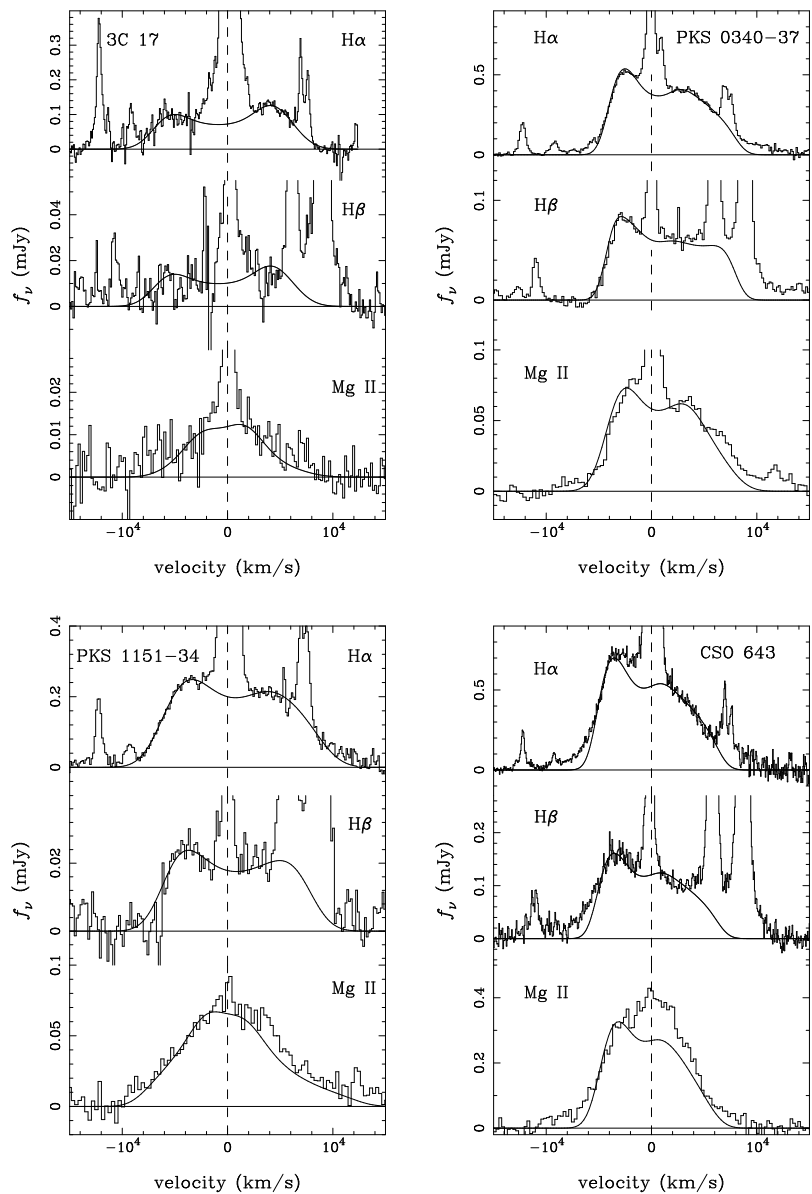}}
\caption{ The broad H\a, H\b, and \ion{Mg}{2}\ profiles of
3C~17, PKS~0340--37, PKS~1151--34, and CSO~643. The profiles are plotted on a
common velocity scale and the best-fitting disk-model profiles are
superposed for comparison. The model parameters are summarized in
Table~\ref{Tfit}. In the case of 3C~17 and CSO~643, the model superposed
on H\b\ is a scaled version of the model that was tuned to fit H\a.
\label{Fprof}}
\end{minipage}
\hskip 0.6truein
\begin{minipage}[t]{4.in}
\epsscale{0.9}
\plotone{f4.eps}
\caption{ Distribution of full widths at half maximum (FWHM) and at
zero intensity (FWZI) of broad H\a\ lines. For each quantity we
compare the distribution in double-peaked emitters (upper panels) and in
other radio-loud AGNs in our collection (lower panels).
\label{Fwidths}}
\end{minipage}
\end{figure}
\begin{figure} 
\hskip -0.6 truein
\begin{minipage}[t]{4in}
\epsscale{1.0}
\centerline{\plotone{f5.eps}}
\caption{ Distribution of fractional wavelength shifts
($\Delta\lambda/\lambda$) of broad H\a\ lines at half maximum and at
zero intensity. In each panel we show the distribution of all objects
in our collection and we identify the double-peaked emitters by shaded
bins.
\label{Fshifts}}
\end{minipage}
\hskip 0.6truein
\begin{minipage}[t]{4in}
\epsscale{0.9}
\centerline{\plotone{f6.eps}}
\caption{ Distributions of (a) percentage starlight in the continuum
around H\a, (b) equivalent with of [\ion{O}{1}]~\l6300 and (c) equivalent
with of [\ion{S}{2}]~\tl6717,~6731.  In each case the distribution in
double-peaked emitters (upper panel) is compared to the distribution in
other radio-loud AGNs from our collection (lower panel).
\label{Fstar}}
\end{minipage}
\end{figure}
}

\begin{figure}   
\epsscale{1.0}
\plotone{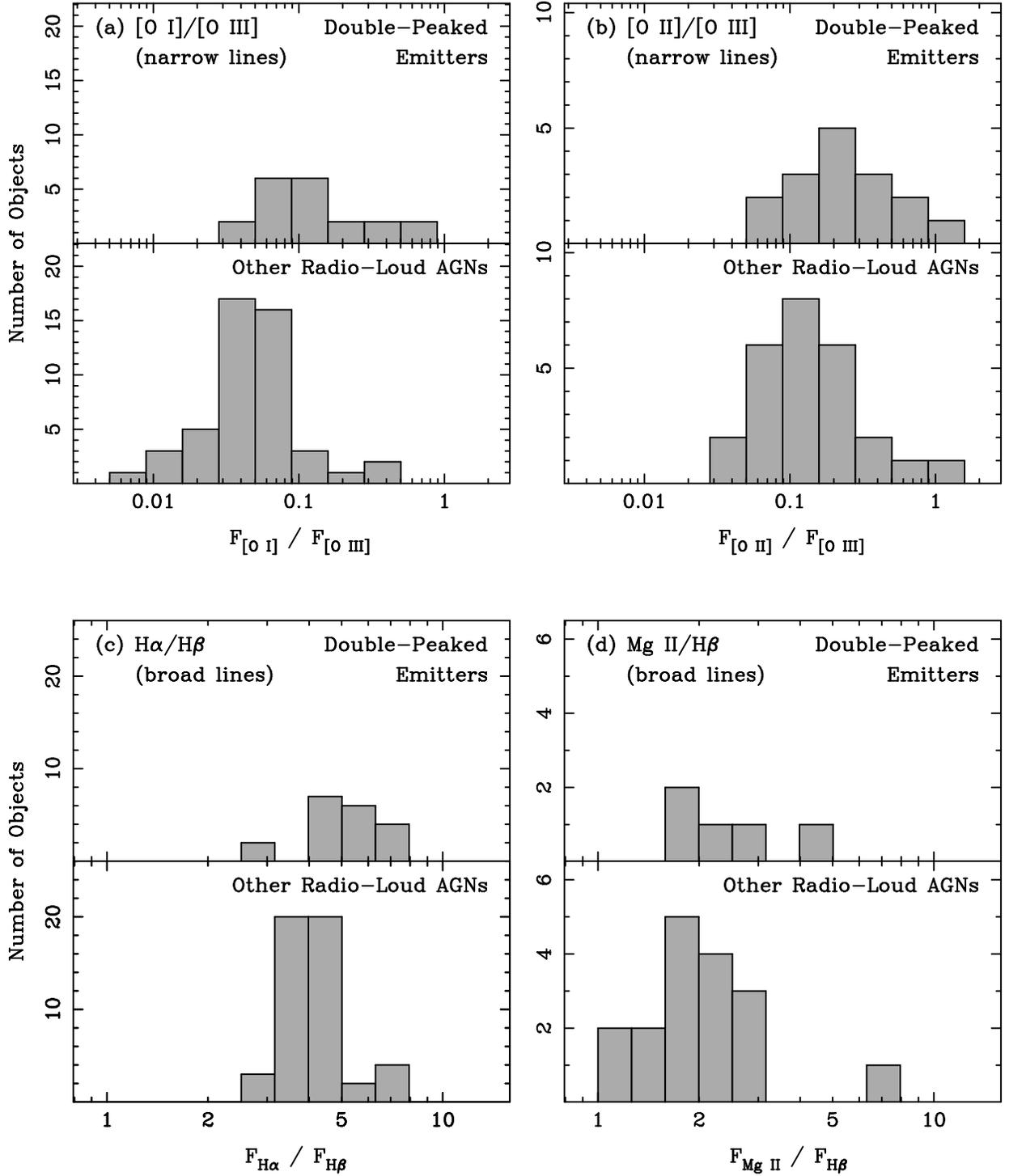}
\caption{ The distributions of the narrow [\ion{O}{1}]/[\ion{O}{3}],
[\ion{O}{2}]/[\ion{O}{3}] ratios and the broad H\a/H\b\ and
\ion{Mg}{2}/H\b\ ratios. The upper panel in each pair shows the
distribution of the corresponding line ratio ratio in double-peaked
emitters while the lower panels shows the distribution of the same
ratio in the comparison sample. All line ratios have been corrected
for Galactic reddening. The values are tabulated in Table~\ref{Tlrat}.
\label{Flrat}}
\end{figure}

\begin{figure}  
\epsscale{0.7}
\plotone{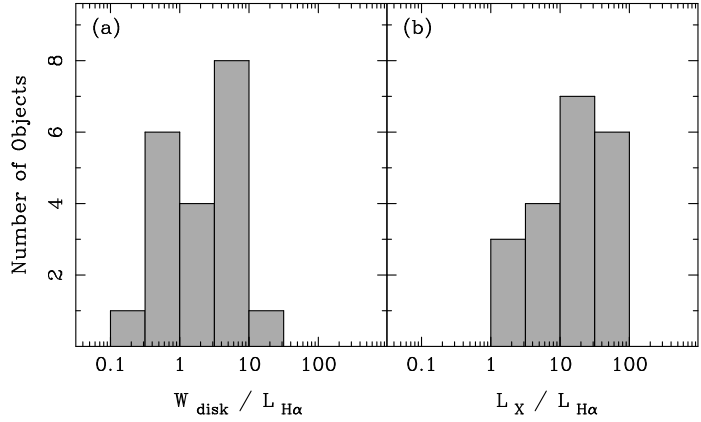}
\caption{
The distribution of (a) the ratio of the accretion power available in
the line-emitting portion of the accretion disk to the H\a\ luminosity
($W_{\rm disk}/L_{\rm H\alpha}$), and (b) the ratio of the X-ray
luminosity to the H\a\ luminosity ($L_{\rm X}/L_{\rm H\alpha})$.
Notice that $W_{\rm disk}/L_{\rm H\alpha} < 1$ in 30\% of the cases
and $W_{\rm disk}/L_{\rm H\alpha} < 10$ in 90\% of the cases.
\label{Fdisk}}
\end{figure}

\begin{figure}  
\epsscale{0.9}
\plotone{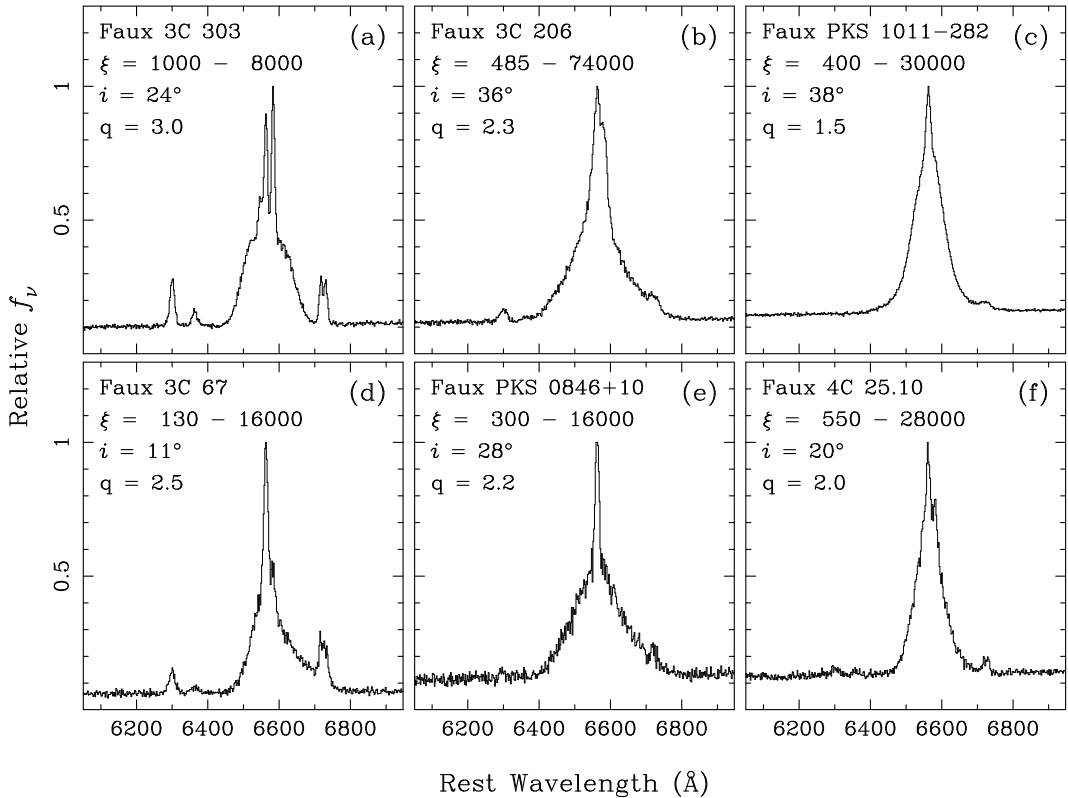}
\caption{ Simulated spectra of the H\a\ region. Disk models with
parameters as indicated are used for the broad H\a\ profiles while the
profiles of the narrow lines are described by Gaussians of FWHM of a
few hundred \kms. To make the spectra look as realistic as possible
Poisson noise has also been added, assuming $S/N\approx 30$--200 in
the continuum. The model parameters were tuned so that these simulated
spectra resemble the observed spectra of objects from paper~I (as
labeled the figure).
\label{Fsimul}}
\end{figure}

\clearpage
\begin{deluxetable}{llllrll}
\tabletypesize{\footnotesize}
\tablewidth{0in}
\tablecolumns{7}
\tablecaption{Journal of Observations\label{Tjour}}
\tablehead{
\colhead{} &
\colhead{} &
\colhead{} &
\colhead{} &
\colhead{Exposure} &
\colhead{} &
\colhead{} \\
\colhead{Object} &
\colhead{$m_{_{\rm V}}$} &
\colhead{$z$} &
\colhead{UT Date} &
\colhead{Time (s)} &
\colhead{Telescope} &
\colhead{Notes} 
}
\startdata
3C 17         \tablenotemark{a} & 18.0 & 0.220 & 1994 Aug  5 &  1800 & Lick 3m & H\a, H\b, \ion{Mg}{2} \\
                                &      &       & 1996 Oct 11 &  3600 & Lick 3m & H\a, H\b, \ion{Mg}{2} \\
MRC 0041+11   \tablenotemark{b} & 19.0 & 0.226 & 1994 Aug 5  &   600 & Lick 3m & \\ 
4C 31.06      \tablenotemark{b} & 18.0 & 0.373 & 1993 Dec 13 &  2000 & KPNO 4m & (B2 0154+31A) \\ 
PKS 0340--37  \tablenotemark{a} & 18.6 & 0.285 & 1994 Feb 15, 17 & $2\times3600$ & CTIO 4m & H\a, H\b, \ion{Mg}{2} \\
3C 93         \tablenotemark{a} & 19.2 & 0.357 & 1993 Dec 13 & 10800 & KPNO 4m & \\
3C 111        \tablenotemark{b} & 18.0 & 0.049 & 1993 Dec 13 &  3200 & KPNO 4m & \\ 
PKS 0511--48  \tablenotemark{b} & 19.0 & 0.306 & 1994 Feb 15 &  4800 & CTIO 4m & narrow lines \\ 
Pictor A      \tablenotemark{{\rm b,c}} & 16.2 & 0.035 & 1994 Feb 17 &  3000 & CTIO 4m & \\ 
CBS 74        \tablenotemark{{\rm b,d}} & 16.0 & 0.092 & 1998 Jan 30 &  3600 & MDM 2.4m & \\
PKS 0921--213 \tablenotemark{{\rm b,d}} & 16.5 & 0.053 & 1995 Mar 24 &  6600 & KPNO 4m & \\ 
4C 72.16      \tablenotemark{b} & 17.9 & 1.462 & 1993 Dec 13 &  2700 & CTIO 4m & new redshift \\ 
4C 36.18      \tablenotemark{a} & 18.0 & 0.392 & 1993 Dec 13 &  7200 & KPNO 4m & \\
PKS 1151--34  \tablenotemark{a} & 17.8 & 0.258 & 1994 Feb 15, 16 &  2700+3600 & CTIO 4m & H\a, H\b, \ion{Mg}{2} \\
PKS 1335--12  \tablenotemark{b} & 18.5 & 0.539 & 1994 Feb 15 &  2700 & CTIO 4m & new redshift \\ 
PKS 1346--11  \tablenotemark{b} & 18.0 & 0.341 & 1994 Feb 15 &  3000 & CTIO 4m & \\ 
CSO 643       \tablenotemark{{\rm b,d}} & 16.7 & 0.276 & 1997 Jun 9  &  2700 & KPNO 2.1m & H\b\ \\
                                &      &       & 1998 Apr 9  &  5400 & MDM 2.4m & H\a\ \\
                                &      &       & 2000 May 31 &  1800 & MDM 2.4m & H\b,\ion{Mg}{2} \\
PKS 1451--37  \tablenotemark{b} & 16.7 & 0.314 & 1994 Feb 15 &  1800 & CTIO 4m & \\ 
PKS 1514+00   \tablenotemark{b} & 15.6 & 0.053 & 1995 Jun 4  &  3600 & KPNO 2.1m &  \\ 
3C 351        \tablenotemark{a} & 15.3 & 0.372 & 1994 Jul 4  &  3600 & KPNO 2.1m & \\
3C 381        \tablenotemark{b} & 17.2 & 0.161 & 1994 Jul 2, 3  & $2\times 2700$ & KPNO 2.1m & narrow lines\\ 
PKS 2247+14   \tablenotemark{b} & 16.9 & 0.235 & 1993 Dec 12 &  1407 & KPNO 4m & \\ 
PKS 2305+18   \tablenotemark{b} & 17.5 & 0.313 & 1993 Dec 12 &  1533 & KPNO 4m & \\ 
PKS 2312--319 \tablenotemark{b} & 18.5 & 1.322 & 1994 Aug 5  &  1800 & Lick 3m & new redshift \\ 
3C 456        \tablenotemark{b} & 18.5 & 0.233 & 1993 Dec 13 &  3600 & KPNO 4m & narrow lines \\ 
MRC 2328+167  \tablenotemark{b} & 18.3 & 0.280 & 1993 Dec 13 &  2400 & KPNO 4m & \\ 
\enddata
\tablenotetext{a}{\small
Repetitions from paper I. See \S2.2 of the text for further details.}
\tablenotetext{b}{\small First-time observations.}
\tablenotetext{c}{\small 
Our observations of Pictor~A were first reported by Halpern \& Eracleous (1994).}
\tablenotetext{d}{\small
Double-peaked emitters originally found by other authors. See \S2.2 of the text for further details.}
\end{deluxetable}
 
\begin{deluxetable}{lrrcrrrr}
\tabletypesize{\footnotesize}
\tablewidth{0in}
\tablecolumns{8}
\tablecaption{Narrow-line EWs, Starlight Fractions, and Broad H$\alpha$ Widths and Shifts\tablenotemark{a}\label{Tstar}}
\tablehead{
\multicolumn{4}{c}{} &
\multicolumn{4}{c}{Broad H\a\ Profile Properties} \\
\noalign{\vskip -4pt}
\multicolumn{4}{c}{} &
\multicolumn{4}{c}{\hrulefill} \\
\colhead{} &
\multicolumn{2}{c}{Obs. EW (\AA)} &
\colhead{} &
\multicolumn{2}{c}{Width (km s$^{-1}$)} &
\multicolumn{2}{c}{$\Delta\lambda/\lambda$} \\
\noalign{\vskip -4pt}
\colhead{} &
\multicolumn{2}{c}{\hrulefill} &
\colhead{Starlight} &
\multicolumn{2}{c}{\hrulefill} &
\multicolumn{2}{c}{\hrulefill} \\
\colhead{} &
\colhead{[\ion{O}{1}]} &
\colhead{[\ion{S}{2}]} &
\colhead{Fraction} &
\colhead{at ZI} &
\colhead{at HM} &
\colhead{at ZI} &
\colhead{at HM} \\
\colhead{Object} &
\colhead{($\pm20$\%)} &
\colhead{($\pm25$\%)} &
\colhead{($\pm0.2$)} &
\colhead{($\pm3000$)} &
\colhead{($\pm200$)} &
\colhead{($\pm0.005$)} &
\colhead{($\pm0.0005$)}
}
\startdata
MC 0041+11   & 11.9 &  4.9 & 0.28 &  8500 &  4200 &   0.0003 & --0.0016 \\
4C 31.06     &  5.4 & 23.3 & 0.11 & 18000 &  9000 &   0.0049 &   0.0001 \\
3C 111       &  7.7 &  6.3 & 0.04 & 18400 &  4800 &   0.0037 &   0.0010 \\
Pictor A     & 55.7 & 30.7 & 0.14 & 29000 & 18400 &   0.0001 & --0.0011 \\
CBS 74       &  7.7 & 14.4 & 0.17 & 22100 &  9200 &   0.0014 &   0.0035 \\
PKS 0921--23 &  5.3 &  6.2 & 0.65 & 21900 &  8300 &   0.0015 & --0.0002 \\
PKS 1346--11 &  4.2 &  5.2 & 0.00 & 11500 &  2300 & --0.0009 &   0.0003 \\
CSO 643      & 10.1 & 26.6 & 0.28 & 21000 &  9000 &   0.0009 & --0.0022 \\
PKS 1451--37 &  2.6 &  4.9 & 0.00 & 15400 &  3800 &   0.0044 &   0.0003 \\
PKS 1514--00 & 10.8 & 15.1 & 0.76 &  8300 &  4300 &   0.0022 &   0.0001 \\
PKS 2247+14  &  0.0 &  4.4 & 0.00 &  8400 &  3500 &   0.0027 &   0.0013 \\
PKS 2305+18  & 14.6 & 15.8 & 0.56 &  8800 &  4400 &   0.0006 &   0.0013 \\
PKS 2328+167 &  0.0 &  6.3 & 0.37 &  8800 &  3200 &   0.0017 &   0.0003 \\
\enddata
\tablenotetext{a}{The uncertainty in the values of each column is 
given under the column heading.}
\end{deluxetable}

%
\begin{deluxetable}{lccc}
\tabletypesize{\footnotesize}
\tablewidth{0in}
\tablecolumns{4}
\tablecaption{Statistical Comparison of Spectroscopic Properties of 
Double-Peaked Emitters and Other Radio-Loud AGNs\label{Tcomp}}
\tablehead{
\colhead{} & \multicolumn{2}{c}{Mean Values} & \colhead{} \\
\noalign{\vskip -4pt}
\colhead{} & \multicolumn{2}{c}{\hrulefill}  & \colhead{} \\
\colhead{Quantitiy} &
\colhead{Double-Peaked Emitters} &
\colhead{Other Radio-Loud AGNs} &
\colhead{K-S Probability \tablenotemark{a}} 
}
\startdata
FWHM of H\a\ line (\kms)          & 12,700 & 62,00  & $2\times 10^{-8}$ \\
FWZI of H\a\ line (\kms)          & 21,700 & 17,000 & $5\times 10^{-3}$ \\
Starlight Fraction                & 0.33  & 0.11 & $3\times 10^{-5}$ \\
Rest EW of [\ion{O}{1}] (\AA)     &  6.8  &  5.0 & $9\times 10^{-2}$ \\
Rest EW of [\ion{S}{2}] (\AA)     & 12.3  &  7.7 & $1\times 10^{-3}$ \\
{[\ion{O}{1}]/[\ion{O}{3}]} Ratio & 0.19  & 0.07 & $7\times 10^{-2}$ \\
{[\ion{O}{2}]/[\ion{O}{3}]} Ratio & 0.35  & 0.21 & $9\times 10^{-2}$ \\
Broad H$\alpha$/H$\beta$ Ratio    & 5.23  & 4.26 & $5\times 10^{-4}$ \\
Broad \ion{Mg}{2}/H$\beta$ Ratio  & 2.62  & 2.19 & $3\times 10^{-1}$  \\
\enddata
\tablenotetext{a}{The probability that the distributions of this quantity
in double-peaked emitters and in other radio-loud AGNs were drawn from the 
same parent population, according to the Kolmogorov-Smirnov test.}
\end{deluxetable}


\begin{deluxetable}{llllcllll}
\tabletypesize{\footnotesize}
\tablewidth{0in}
\tablecolumns{7}
\tablecaption{Disk Profile Parameters\label{Tfit}}
\tablehead{
\colhead{} &
\colhead{} &
\colhead{} &
\colhead{$i$} &
\colhead{} &
\colhead{} &
\colhead{$\sigma$} &
\colhead{$\varphi_0$} &
\colhead{} \\
\colhead{Object} &
\colhead{$\xi_1$} &
\colhead{$\xi_2$} &
\colhead{(\deg)} &
\colhead{$q^{\;\rm a}$} &
\colhead{$\xi_{\rm b}^{\;\rm a}$} &
\colhead{(\kms)} &
\colhead{(\deg)} &
\colhead{$e$} 
}
\startdata
\sidehead{\underbar{H\a\ \it Profiles}}
3C 17 (elliptical) $^{\;\rm b}$ & 360$^{+40}_{-110}$ & 800$\pm$200           & 30$\pm$4       & 1.7 & \dots & 1400$\pm$200 & 100$^{+20}_{-30}$ & 0.3$\pm$0.1 \\
4C 31.06                  & $>1000$            & $>14000$              & $>35$          & 3.0   & \dots & 1200$\pm$200 & \dots & \dots \\
PKS 0340--37$^{\;\rm c}$  & 210$^{+50}_{-30}$  & 1900$\pm$600          & 18$\pm$1       & 3.0   & \dots & 1400$^{+200}_{-300}$  & \dots & \dots \\ 
Pictor A (circular) $^{\;\rm d}$   & 1000$^{+130}_{-250}$  & 1400$^{+200}_{-300}$ & 70$^{+20}_{-15}$ & 3.0   & \dots & 1000$\pm$200 & \dots & \dots \\
Pictor A (elliptical) $^{\;\rm b,d}$ &    220$^{+50}_{-20}$  &  550$^{+70}_{-30}$   & 30$^{+3}_{-1}$   & 1.5   & \dots & 1000$\pm$200 & 110$^{+15}_{-50}$ & 0.29$_{-0.06}^{+0.26}$ \\
CBS 74                    & 170$^{+60}_{-20}$  & 1800$^{+700}_{-100}$  & 16$^{+3}_{-1}$ & 3.0   & \dots & 1000$^{+500}_{-200}$  & \dots & \dots \\ 
PKS 0921--213             & $>850$             & $>5600$               & $>40$          & 1.5   & \dots &  600$\pm$100 & \dots & \dots \\
PKS 1151--34$^{\;\rm e}$  & 300$^{+170}_{-40}$ & 2000$^{+1100}_{-200}$ & 28$^{+8}_{-2}$ & 3.0   & \dots & 1600$^{+200}_{-400}$ & \dots & \dots \\
CSO 643 (blueshifted)$^{\;\rm f}$    & 210$^{+80}_{-50}$  & 2700$^{+1000}_{-600}$ & 17$^{+5}_{-2}$ & 3.0   & \dots & 1000$\pm$200 & \dots & \dots \\
CSO 643 (elliptical)$^{\;\rm b,f}$     & 1600$\pm$800  & 10000$\pm$5000 & 55$^{+35}_{-20}$ & 3.0   & \dots & 1200$\pm$200 & 90$\pm$40 & 0.5$_{-0.1}^{+0.2}$\\
\sidehead{\underbar{H\b\ \it Profiles}}
3C 17 (elliptical) $^{\;\rm b}$ & \multicolumn{8}{l}{same parameters as H\a\ profile} \\
PKS 0340--37              & 220$^{+30}_{-40}$  & $>1500$                & \dots      & 5.1, 3.0 & 450$\pm$100          & \dots & \dots & \dots \\ 
PKS 1151--34              & 400$^{+100}_{-50}$ & 2000$\pm$500           & \dots      & 5.1, 3.0 & 1000$^{+100}_{-600}$ &\dots & \dots & \dots \\
CSO 643 (blueshifted)$^{\;\rm f}$ & \multicolumn{8}{l}{same parameters as H\a\ profile} \\
\sidehead{\underbar{\ion{Mg}{2} \it Profiles}}
3C 17 (elliptical)        & \multicolumn{8}{l}{same parameters as H\a\ profile but with $\xi_2 > 5000$} \\
PKS 0340--37              & 200$^{+50}_{-20}$  & 2500$^{+500}_{-600}$   & \dots      & 2.2, 3.0 & 700$^{+200}_{-100}$  &\dots & \dots & \dots \\ 
PKS 1151--34              & 170$^{+30}_{-20}$  & $>15000$               & \dots      & 2.2, 3.0 & 3000$\pm$1000        & \dots & \dots & \dots \\
CSO 643 (blueshifted)$^{\;\rm e}$ & 200$^{+30}_{-40}$  & 3000$\pm$1000  & \dots      & 2.2, 3.0 & 500$^{+200}_{-100}$  & \dots  & \dots & \dots \\
\enddata
\tablenotetext{a} 
{A single value of the emissivity power law index is used for H\a\
profiles. For H\b\ and \ion{Mg}{2} profiles two values of the power
law index are given, the first corresponding to the inner disk and the
second to the outer disk; the transition occurs at the break radius
$\xi_{\rm b}$. The power-law index is generally fixed at the values
predicted by photionization models, with the exception of the H\a\
profiles of 3C~17 and Pictor~A (elliptical disk models), and
PKS~0921--213. See the discussion in \S3 of the text.}
\tablenotetext{b}
{In the case of elliptical disk models, $\xi_1$ and $\xi_2$ are the
inner and outer pericenter distances.  The eccentricity of the disk
increases linearly with radius, from 0 to $e$, while the major axis is
oriented at an angle $\varphi_0$ to the line of sight.}
\tablenotetext{c}
{Best-fitting model parameters taken from paper~I and quoted here for
reference.}
\tablenotetext{d} 
{The H\a\ profile of Pictor A was fitted with two different models:
one of a circular disk and one of an elliptical disk (see also \S3).}
\tablenotetext{e} 
{Parameters taken from paper~I, but adjusted slightly to accommodate a
small variation of the profile since the earlier observation.  The
main change is in the value of $\sigma$.}
\tablenotetext{f} 
{The H\a\ profile of CSO 643 was fitted with two different models: one
of a circular disk and one of an elliptical disk (see also \S3).}
\end{deluxetable}

\clearpage
\begin{deluxetable}{llllllll}
\tabletypesize{\footnotesize}
\tablewidth{0in}
\tablecolumns{8}
\tablecaption{Emission-Line Ratios\tablenotemark{a}\label{Tlrat}}
\tablehead{
\colhead{} &
\multicolumn{2}{c}{\hbox to 7em{\hfil Narrow Lines\hfil}} &
\colhead{\hbox to 1 em{\hss}} &
\multicolumn{2}{c}{\hbox to 6em{\hfil Broad Lines\hfil}} &
\colhead{\hbox to 1 em{\hss}} &
\colhead{} \\
\noalign{\vskip -6pt}
\colhead{} &
\multicolumn{2}{c}{\hbox to 7em{\hrulefill}} &
\colhead{\hbox to 1 em{\hss}} &
\multicolumn{2}{c}{\hbox to 6em{\hrulefill}} &
\colhead{\hbox to 1 em{\hss}} &
\colhead{} \\
\colhead{} &
\colhead{[\ion{O}{1}]} &
\colhead{[\ion{O}{2}]} &
\colhead{\hbox to 1 em{\hss}} &
\colhead{H\a} &
\colhead{\ion{Mg}{2}} &
\colhead{\hbox to 1 em{\hss}} &
\colhead{} \\
\noalign{\vskip -4pt}
\colhead{} &
\colhead{\hbox to 3 em{\hrulefill}} &
\colhead{\hbox to 3 em{\hrulefill}} &
\colhead{\hbox to 1 em{\hss}} &
\colhead{\hbox to 1.8 em{\hrulefill}} &
\colhead{\hbox to 2.5 em{\hrulefill}} &
\colhead{\hbox to 1 em{\hss}} &
\colhead{} \\
\colhead{\hbox to 8em{\hfil Object\hfil}} &
\colhead{[\ion{O}{3}]} &
\colhead{[\ion{O}{3}]} &
\colhead{\hbox to 1 em{\hss}} &
\colhead{H\b} &
\colhead{H\b} &
\colhead{\hbox to 1 em{\hss}} &
\colhead{\hbox to 8em{References\tablenotemark{b}\hfil}}
}
\startdata
\multicolumn{7}{c}{\hfil Double-Peaked Emitters\hfil} \\
\noalign{\vskip 6pt \hrule \vskip 6pt}
3C 17             & 0.43  & 0.74   & &  7.19\tablenotemark{\;{\rm c}} &  1.34\tablenotemark{\;{\rm c}} & & 1          \\ 
4C 31.06          & 0.048 & 0.40   & &  4.90 & \dots & & 1          \\ 
3C 59             & 0.066 & 0.15   & &  4.64 & \dots & & 3          \\ 
IRAS 0236.6--3101 & 0.10  & \dots  & & \dots & \dots & & 1,2        \\ 
PKS 0340--37      & 0.083 & 0.27   & &  5.28 &  1.91 & & 1          \\ 
3C 93             & 0.055 & 0.26   & &  4.38 & \dots & & 1          \\ 
MS 0450.3--1817   & 0.46  & 0.70   & & \dots & \dots & & 1          \\ 
Pictor A          & 0.58  & 0.50   & &  6.65 & \dots & & 1          \\ 
B2 0742+31        & 0.028 & \dots  & &  5.16 & \dots & & 1          \\ 
CBS 74            & 0.087 & 0.18   & &  3.00 & \dots & & 1          \\ 
PKS 0857--19      & 0.16  & \dots  & &  4.78 & \dots & & 1          \\ 
PKS 0921--213     & 0.22  & 0.11   & &  4.59 & \dots & & 1          \\ 
4C 36.18          & 0.11  & 0.22   & &  5.63 & \dots & & 1          \\ 
PKS 1151--34      & 0.10  & 0.33   & &  7.32 &  4.06 & & 1          \\ 
CSO 643           & \dots & 0.071  & &  3.00 &  3.15 & & 1          \\ 
3C 332            & 0.14  & 0.16   & &  7.38 & \dots & & 3          \\ 
Arp 102B          & 0.78  & 1.2    & &  4.41 &  1.83 & & 4          \\ 
PKS 1739+18C      & 0.15  & \dots  & &  5.27 & \dots & & 1          \\ 
3C 382            & 0.070 & 0.15   & &  4.82 & \dots & & 5          \\ 
3C 390.3          & 0.064 & 0.070  & &  5.84 & \dots & & 5          \\ 
PKS 1914--45      & 0.12  & \dots  & &  5.10 & \dots & & 1          \\ 
\noalign{\vskip 6pt \hrule \vskip 6pt}
\multicolumn{7}{c}{\hfil Other Radio-Loud AGNs \hfil} \\
\noalign{\vskip 6pt \hrule \vskip 6pt}
3C 18             & \dots & 0.16   & & \dots & \dots & & 10         \\ 
MRC 0041+11       & 0.13  & 0.14   & &  5.20 &  1.43 & & 1          \\ 
TXS 0042+101      & 0.42  & \dots  & & \dots & \dots & & 8          \\ 
3C 48             & \dots & 0.17   & &  6.76 &  1.10 & & 6,7        \\ 
3C 47             & \dots & 0.13   & & \dots &  2.50 & & 8,9        \\ 
PKS 0202--76      & 0.053 & \dots  & &  4.21 & \dots & & 1          \\ 
PKS 0214+10       & 0.039 & \dots  & &  4.90 & \dots & & 1          \\ 
3C 67             & \dots & 0.31   & & \dots & \dots & & 11         \\ 
PKS 0403--132     & \dots & 0.20   & & \dots &  1.70 & & 8,9        \\ 
3C 111            & 0.076 & \dots  & &  2.94 & \dots & & 1          \\ 
3C 120            & 0.059 & 0.047  & &  4.98\tablenotemark{\;{\rm d}} & \dots & & 10,12    \\ 
3C 147            & \dots & 0.46   & & \dots & \dots & & 8          \\ 
PKS 0736+01       & \dots & 0.093  & &  3.37\tablenotemark{\;{\rm d}} &  1.78 & & 10,6,13  \\ 
3C 206            & \dots & 0.097  & &  4.73 & \dots & & 8,14       \\ 
4C 05.38          & 0.048 & \dots  & & \dots & \dots & & 1          \\ 
3C 215            & \dots & 0.17   & & \dots &  1.40 & & 15,9       \\ 
PKS 0925--203     & 0.038 & \dots  & &  4.69 & \dots & & 1          \\ 
3C 227            & 0.072 & 0.081  & &  6.37 & \dots & & 5          \\ 
4C 09.35          & 0.035 & \dots  & & \dots & \dots & & 1          \\ 
3C 232            & \dots & 0.32   & & \dots &  7.74\tablenotemark{\;{\rm d}} & & 13       \\ 
3C 234            & 0.013 & 0.075  & & \dots & \dots & & 3,6        \\ 
PKS 1004+13       & \dots & \dots  & &  4.14 & \dots & & 16         \\ 
B2 1028+31        & 0.031 & \dots  & &  3.85 & \dots & & 8,6        \\ 
3C 246            & 0.042 & \dots  & &  4.87 & \dots & & 1          \\ 
4C 61.20          & \dots & 0.30   & & \dots & \dots & & 13         \\ 
3C 249.1          & \dots & \dots  & &  3.91 & \dots & & 16         \\ 
PKS 1101--32      & 0.044 & \dots  & &  4.20 & \dots & & 1          \\ 
PKS 1103--006     & \dots & \dots  & & \dots &  1.60 & & 9          \\ 
PKS 1136--135     & \dots & 0.083  & & \dots & \dots & & 10         \\ 
ON 029            & \dots & \dots  & &  3.73 & \dots & & 8          \\ 
B2 1223+25        & 0.020 & \dots  & &  3.54 & \dots & & 1          \\ 
3C 273            & \dots & 0.16   & &  4.26\tablenotemark{\;{\rm d}} &  0.35 & & 10,13    \\ 
PKS 1233--24      & 0.016 & \dots  & &  5.68 & \dots & & 1          \\ 
3C 277.1          & 0.35  & 0.081  & &  3.90 &  2.30 & & 8,11,9     \\ 
PKS 1302--102     & 0.080 & \dots  & &  3.18 & \dots & & 16         \\ 
3C 287.1          & 0.24  & 0.74   & & \dots & \dots & & 3          \\ 
PKS 1346--11      & 0.054 & \dots  & &  3.07 & \dots & & 1          \\ 
B2 1351+26        & 0.061 & \dots  & &  3.56 & \dots & & 16         \\ 
PKS 1355--41      & 0.13  & 0.13   & &  4.53 & \dots & & 1,10       \\ 
Mkn 668           & 0.036 & \dots  & & \dots & \dots & & 1          \\ 
PKS 1417--19      & 0.033 & 0.046  & &  4.83 & \dots & & 3          \\ 
PKS 1421--38      & 0.066 & \dots  & &  3.54 & \dots & & 1          \\ 
B2 1425+26        & 0.074 & \dots  & &  6.64 & \dots & & 16         \\ 
PKS 1451--37      & 0.040 & \dots  & &  4.04 & \dots & & 1          \\ 
4C 69.18          & 0.063 & \dots  & &  4.14 & \dots & & 16         \\ 
PKS 1510--08      & \dots & 0.17   & & \dots &  2.23 & & 1,10,6     \\ 
4C 37.43          & 0.028 & \dots  & &  4.21 & \dots & & 1          \\ 
PKS 1514+00       & 0.50  & 1.2    & & \dots & \dots & & 1          \\ 
LB 9743           & 0.007 & \dots  & &  4.19 & \dots & & 16         \\ 
3C 323.1          & 0.025 & \dots  & &  2.98 &  1.80 & & 16,9       \\ 
3C 334            & \dots & \dots  & & \dots &  2.70 & & 9          \\ 
3C 345            & \dots & \dots  & & \dots &  3.00 & & 9          \\ 
4C 18.47          & \dots & \dots  & &  3.75 & \dots & & 16         \\ 
4C 61.34          & \dots & 0.69   & & \dots & \dots & & 8          \\ 
3C 351            & 0.015 & 0.087  & & \dots &  1.97 & & 6          \\ 
B2 1719+35        & 0.059 & \dots  & &  3.45 & \dots & & 1,16       \\ 
4C 34.47          & 0.070 & 0.058  & &  4.44 &  2.58 & & 6          \\ 
PKS 1725+044      & 0.030 & \dots  & &  3.42 & \dots & & 16         \\ 
MRC 1745+16       & 0.043 & \dots  & &  3.93 & \dots & & 1          \\ 
3C 381            & 0.053 & 0.23   & & \dots & \dots & & 3          \\ 
4C 73.18          & 0.030 & \dots  & &  3.87 & \dots & & 8,6        \\ 
PKS 2128--12      & \dots & 0.088  & & \dots & \dots & & 10         \\ 
PKS 2135--14      & 0.092 & 0.090  & &  4.48 & \dots & & 10,16      \\ 
PKS 2140--048     & 0.054 & \dots  & &  3.33 & \dots & & 1          \\ 
OX 169            & 0.025 & \dots  & &  3.23 & \dots & & 1,8        \\ 
4C 31.63          & \dots & \dots  & &  3.48 &  2.30 & & 8,9        \\ 
PKS 2208--13      & 0.071 & \dots  & &  3.90 & \dots & & 1          \\ 
3C 445            & 0.042 & 0.078  & &  6.63 & \dots & & 5          \\ 
PKS 2227--399     & 0.058 & \dots  & & \dots & \dots & & 1          \\ 
PKS 2247+14       & 0.018 & \dots  & &  5.00 & \dots & & 1          \\ 
4C 11.72          & \dots & 0.13   & &  4.90 &  1.00 & & 8,9        \\ 
PKS 2302--71      & 0.046 & \dots  & &  3.93 & \dots & & 1          \\ 
PKS 2305+18       & 0.046 & \dots  & &  3.98 & \dots & & 1,16       \\ 
PKS 2345--167     & \dots & 0.20   & & \dots & \dots & & 8          \\ 
PKS 2349--01      & 0.038 & 0.28   & &  4.07 & \dots & & 3          \\ 
\enddata
\tablenotetext{a}{\small
All objects have redshifts less than 0.6.  Line ratios have been
corrected for Galactic reddening using the Seaton (1979) law and the
color excess reported by Schlegel et al. (1998). Uncertainties on line
ratios from this work and from paper~I are typically 20\%.}
\tablenotetext{b}{\small
{\it References. -- }
 (1)  Eracleous \& Halpern 1994, and this work,
 (2)  Colina et al. 1991,                          
 (3)  Grandi \& Osterbrock  1978,                  
 (4)  Halpern et al. 1996,                         
 (5)  Osterbrock et al. 1976,                      
 (6)  Grandi \& Phillips 1979,                     
 (7)  Thuan et al. 1979,                           
 (8)  Jackson \& Browne 1991a,                      
 (9)  Wills et al. 1995; Brotherton, M. S. \& Wills, B. J. private communication,                      
 (10) Tadhunter et al. 1993,                       
 (11) Gelderman, \& Whittle 1994,                  
 (12) Osterbrock, 1977,                            
 (13) Phillips 1978,                               
 (14) Baldwin 1975,                                
 (15) Wills et al. 1993,                           
 (16) Jackson \& Eracleous 1995.}
\tablenotetext{c}{\small
The H\a/H\b\ and \ion{Mg}{2}/H\b\ ratios of 3C~17 are fairly uncertain
because of the weakness of H\b. The H\a/\ion{Mg}{2} ratio, however, is
not as uncertain; its value is 5.36.}  
\tablenotetext{d}{\small
The reported line ratio refers to the total flux of the broad and
narrow lines combined.}
\end{deluxetable}

\clearpage
\begin{deluxetable}{lcrll}
\tabletypesize{\footnotesize}
\tablewidth{0in}
\tablecolumns{5}
\tablecaption{Radio Properties of Double-Peaked Emitters\label{Tradio}}
\tablehead{
\colhead{} &
\colhead{$L_{\rm \;5~GHz}$} &
\colhead{} &
\colhead{} &
\colhead{} \\
\colhead{Object} &
\colhead{(W Hz\m1)} &
\colhead{$s$\tablenotemark{a}} &
\colhead{Morphology} &
\colhead{Refs\tablenotemark{b}} 
}
\startdata
3C 17             & $6.5\times 10^{26}$ & 0.62 & Fragmented lobes + core & 1 \cr
4C 31.06          & $2.8\times 10^{26}$ & 0.70 & Extended to 90~kpc (presumably lobes) & 2,3,4 \cr
3C 59             & $4.7\times 10^{25}$ & 0.72 & Twin lobes + compact core & 5 \cr
PKS 0235+023      & $3.8\times 10^{25}$ & 0.72 & Twin lobes + compact core & 6 \cr
IRAS 0236.6--3101 & $6.0\times 10^{22}$ & \dots& \dots   & 7 \cr
PKS 0340--37      & $2.8\times 10^{26}$ & 0.87 & Core with extension (low-res. map) & 8 \cr
3C 93             & $6.0\times 10^{26}$ & 0.93 & Twin lobes & 9 \cr
MS 0450.3--1817   & $4.1\times 10^{23}$ & \dots& Possible lobes + core & 10,11 \cr
Pictor A          & $8.5\times 10^{25}$ & 1.07 & Edge-brightened double lobes & 1,12 \cr
B2 0742+31        & $1.1\times 10^{27}$ & 0.82 & Twin lobes + compact core & 3,14 \cr
CBS 74            & $1.3\times 10^{25}$\tablenotemark{\;c} & \dots & Core + possible lobes & 8,13,14 \cr
PKS 0857--19      & $2.7\times 10^{26}$ & 0.80 & \dots & 15 \cr
PKS 0921--213     & $4.4\times 10^{24}$ & 0.66 & Twin lobes + compact core & 8,15,16 \cr
PKS 1020--103     & $2.0\times 10^{25}$ & 0.92 & Faint lobes + compact core & 17 \cr
4C 36.18          & $1.7\times 10^{26}$ & 0.95 & Twin lobes & 3,18 \cr
PKS 1151--34      & $9.5\times 10^{26}$ &   0.76 & Compact & 1,19 \cr
TXS 1156+213      & $5.0\times 10^{25}$ & 0.78 & \dots & 3 \cr
CSO 643           & $8.3\times 10^{25}$ & 0.62 & Asymmetric lobes + compact core & 8,14 \cr
3C 303            & $8.9\times 10^{25}$ & 0.81 & Twin lobes + compact core & 20 \cr
3C 332            & $9.0\times 10^{25}$ & 0.92 & Twin lobes + compact core & 5 \cr
Arp 102B          & $2.4\times 10^{23}$ & 0.22 & Compact with extended halo & 21 \cr
PKS 1739+18C      & $5.6\times 10^{25}$ & 0.83 & Twin lobes + faint core & 3,8,17 \cr
3C 382            & $3.4\times 10^{25}$ & 0.67 & Twin lobes + compact core & 5 \cr
3C 390.3          & $6.3\times 10^{25}$ & 0.62 & Twin lobes + compact core & 20 \cr
PKS 1914--45      & $2.3\times 10^{26}$ & 0.94 & \dots & 22 \cr
PKS 2300--18      & $5.8\times 10^{25}$ & 0.77 & Asymmetric lobes + compact core & 15,19 \cr
\enddata
\tablenotetext{a}{\small
The spectral index, assuming that $f_{\nu}\propto\nu^{-s}$.}
\tablenotetext{b}{\small {\it References . --} 
 (1) Morganti, Killeen, \& Tadhunter (1993);
 (2) Potash \& Wardle (1979);
 (3) Becker et al. (1991);
 (4) Slee et al. (1994);
 (5) Antonucci (1985);
 (6) Dunlop et al. (1989);
 (7) Roy et al. (1994);
 (8) Condon et al. (1998);
 (9) Hintzen et al (1983);
(10) Feigelson et al. (1982);
(11) Eracleous \& Halpern (1994);
(12) Jones \& McAdam (1992);
(13) Bauer et al. (2000);
(14) Becker et al. (1995);
(15) Griffith et al. (1994);
(16) Condon et al. (1977);
(17) Lister et al. (1994);
(18) Allington-Smith (1982);
(19) Reid et al. (1999);
(20) Leahy \& Perley (1991);
(21) Puschell et al. (1986);
(22) Wright et al. (1994).
}
\tablenotetext{c}{\small
Extrapolated from a measurement at 1.4 GHz using a ``canonical''
spectral index of $s=0.7 $.}
\end{deluxetable}

\clearpage
\begin{deluxetable}{lllllll}
\tabletypesize{\footnotesize}
\tablewidth{0in}
\tablecolumns{7}
\tablecaption{Power Output of Line-Emitting Disks\label{Tdisk}}
\tablehead{
\colhead{} &
\colhead{$F_{\;\rm X}$} &
\colhead{Band} &
\colhead{$L_{\;\rm X}$} &
\colhead{$W_{\;\rm disk}$\tablenotemark{a}} &
\colhead{$L_{\;\rm H\alpha}$} &
\colhead{} \\
\colhead{Object} &
\colhead{(erg cm\m2 s\m1)} &
\colhead{(keV)} &
\colhead{(erg s\m1)} &
\colhead{(erg s\m1)} &
\colhead{(erg s\m1)} &
\colhead{Ref.\tablenotemark{b}} 
}
\startdata
3C 17             & $4.4\times 10^{-13}$ & 0.1--2.4 & $5.2\times 10^{43}$ & $3.4\times 10^{42}$ & $2.6\times 10^{43}$ & 1 \\
4C 31.06          & $9.2\times 10^{-13}$ & 0.1--2.4 & $3.4\times 10^{44}$ & $2.3\times 10^{43}$ & $9.0\times 10^{42}$ & 1 \\
3C 59             & $6.5\times 10^{-12}$ & 0.1--2.4 & $1.8\times 10^{44}$ & $4.4\times 10^{43}$ & $5.9\times 10^{42}$ & 2 \\
PKS 0235+023      & $3.9\times 10^{-12}$ & 0.1--2.4 & $4.0\times 10^{44}$ & $2.5\times 10^{43}$ & $5.9\times 10^{43}$ & 2 \\
IRAS 0236.6--3101 & $1.1\times 10^{-12}$ & 0.1--2.4 & $9.7\times 10^{42}$ & $5.0\times 10^{41}$ & $1.5\times 10^{42}$ & 3 \\
PKS 0340--37      & $1.7\times 10^{-12}$ & 0.1--2.4 & $3.5\times 10^{44}$ & $9.9\times 10^{43}$ & $1.3\times 10^{44}$ & 2 \\
3C 93             & $8.6\times 10^{-13}$ & 0.3--3.5 & $2.9\times 10^{44}$ & $6.0\times 10^{43}$ & $1.2\times 10^{43}$ & 4 \\
MS 0450.3--1817   & $5.4\times 10^{-13}$ & 0.3--3.5 & $4.6\times 10^{42}$ & $2.4\times 10^{41}$ & $1.8\times 10^{41}$ & 5 \\
Pictor A          & $2.1\times 10^{-11}$ & 0.1--2.4 & $5.6\times 10^{43}$ & $3.8\times 10^{42}$ & $5.7\times 10^{41}$ & 6 \\
B2 0742+31        & $3.2\times 10^{-12}$ & 0.1--2.4 & $1.9\times 10^{45}$ & $2.6\times 10^{44}$ & $3.0\times 10^{44}$ & 2 \\
CBS 74            & $9.3\times 10^{-12}$ & 0.1--2.4 & $1.8\times 10^{44}$ & $6.4\times 10^{43}$ & $7.6\times 10^{42}$ & 7 \\
PKS 0921--213     & $7.5\times 10^{-12}$ & 0.1--2.4 & $4.8\times 10^{43}$ & $3.4\times 10^{42}$ & $2.5\times 10^{42}$ & 8 \\
PKS 1020--103     & $6.5\times 10^{-12}$ & 0.1--2.4 & $6.0\times 10^{44}$ & $1.1\times 10^{44}$ & $2.9\times 10^{43}$ & 2 \\
4C 36.18          & $8.4\times 10^{-13}$ & 0.1--2.4 & $3.4\times 10^{44}$ & \dots & \dots & 1  \\
CSO 643           & $4.6\times 10^{-12}$ & 0.1--2.4 & $8.8\times 10^{44}$ & $2.6\times 10^{44}$ & $1.4\times 10^{43}$ & 2 \\
3C 303            & $1.2\times 10^{-12}$ & 0.5--3.0 & $5.6\times 10^{43}$ & $3.7\times 10^{42}$ & $1.7\times 10^{42}$ & 9 \\
3C 332            & $1.0\times 10^{-12}$ & 0.5--3.5 & $5.4\times 10^{43}$ & $1.3\times 10^{43}$ & $2.5\times 10^{42}$ & 10 \\
Arp 102B          & $2.1\times 10^{-12}$ & 0.4--3.8 & $2.8\times 10^{42}$ & $3.6\times 10^{41}$ & $7.3\times 10^{41}$ & 11 \\
PKS 1739+18C      & $6.6\times 10^{-12}$ & 0.1--2.4 & $5.5\times 10^{44}$ & \dots & \dots &  2 \\
3C 382            & $7.5\times 10^{-11}$ & 0.1--2.4 & $5.7\times 10^{44}$ & $5.2\times 10^{43}$ & $1.1\times 10^{43}$ & 2 \\
3C 390.3          & $1.8\times 10^{-11}$ & 0.5--3.0 & $1.3\times 10^{44}$ & $1.6\times 10^{43}$ & $3.5\times 10^{42}$ & 9 \\
PKS 1914--45      & $3.7\times 10^{-13}$ & 0.1--2.4 & $1.3\times 10^{44}$ & $1.8\times 10^{43}$ & $4.2\times 10^{43}$ & 12 \\
PKS 2300--18      & $7.2\times 10^{-12}$ & 0.1--2.4 & $2.8\times 10^{44}$ & \dots & \dots &  8 \\
\enddata
\tablenotetext{a}{\small 
The viscous power output of the line-emitting portion of the accretion disk.}
\tablenotetext{b}{\small {\it References. --} 
 (1) Brinkmann et al. (1995); 
 (2) Bauer et al. (2000);
 (3) Boller et al (1992);
 (4) Eracleous \& Halpern (1994);
 (5) Stocke et al. (1983);
 (6) Brinkmann \& Siebert (1994);
 (7) Brinkmann et al. (2000);
 (8) Siebert et al. (1998);
 (9) Fabbiano et al. (1984);
(10) Halpern (1990);
(11) Biermann et al. (1981);
(12) Voges et al. (1994).
}
\end{deluxetable}

\end{document}